\documentclass[sigconf]{acmart}

\AtBeginDocument{%
  }

\usepackage{tabularx}
\usepackage{graphicx}
\usepackage{subcaption}
\usepackage{multirow}
\usepackage{enumitem}
\usepackage{appendix}
\usepackage{makecell}
\usepackage{needspace}
\usepackage{colortbl}

\newcommand{\sigtwo}[1]{\textcolor{red!60!black}{\underline{#1}}}
\newcommand{\best}[1]{\textcolor{red!60!black}{\underline{#1}}}

\usepackage{graphicx}

\copyrightyear{2026}
\acmYear{2026}
\setcopyright{cc}
\setcctype{by-nc-nd}
\acmConference[UMAP '26]{34th ACM Conference on User Modeling, Adaptation and Personalization}{June 08--11, 2026}{Gothenburg, Sweden}
\acmBooktitle{34th ACM Conference on User Modeling, Adaptation and Personalization (UMAP '26), June 08--11, 2026, Gothenburg, Sweden}
\acmDOI{10.1145/3774935.3806190}
\acmISBN{979-8-4007-2311-7/2026/06}

\begin{document}


\title{Eyes Can’t Always Tell: Fusing Eye Tracking and User Priors for User Modeling under AI Advice Conditions}




\author{Xin Sun}
\affiliation{
  \institution{National Institute of Informatics}
  \city{Tokyo}
  \country{Japan}
}
\email{xsun@nii.ac.jp}

\author{Shu Wei}
\affiliation{%
  \institution{Yale School of Medicine}
  \city{New Haven}
  \country{USA}
}
\email{s.wei@yale.edu}

\author{Ting Pan}
\affiliation{%
  \institution{University of Amsterdam}
  \city{Amsterdam}
  \country{Netherlands}
}
\email{t.pan@uva.nl}

\author{Yajing Wang}
\affiliation{%
  \institution{University of Amsterdam}
  \city{Amsterdam}
  \country{Netherlands}
}
\email{y.wang16@uva.nl}

\author{Jos A Bosch}
\affiliation{%
  \institution{University of Amsterdam}
  \city{Amsterdam}
  \country{Netherlands}
}
\email{j.a.bosch@uva.nl}

\author{Isao Echizen}
\affiliation{%
  \institution{National Institute of Informatics}
  \city{Tokyo}
  \country{Japan}
}
\email{iechizen@nii.ac.jp}

\author{Abdallah El Ali}
\affiliation{%
  \institution{Centrum Wiskunde \& Informatica}
  \country{Netherlands}
}
\affiliation{%
  \institution{Utrecht University}
  \city{Utrecht}
  \country{Netherlands}
}
\email{Abdallah.El.Ali@cwi.nl}

\author{Saku Sugawara}
\affiliation{%
  \institution{National Institute of Informatics}
  \city{Tokyo}
  \country{Japan}
}
\email{saku@nii.ac.jp}




\begin{abstract}

Modeling users’ cognitive states (e.g., cognitive load and decision confidence) is essential for building adaptive AI in high-stakes decision-making. 
While eye tracking provides non-invasive behavioral signals correlated with cognitive effort, prior work has not systematically examined how AI assistance contexts, specifically varying advice reliability and user heterogeneity, can alter the mapping between gaze signals and cognitive states. 
We conducted a within-subject lab eye-tracking study (N=54) on factual verification tasks under three conditions: No-AI, Correct-AI advice, and Incorrect-AI advice. 
We analyze condition-dependent changes in self-reports and eye-tracking patterns and evaluate the robustness of eye-tracking-based user modeling. 
Results show that AI advice increases decision confidence compared to No-AI, while Correct-AI is associated with lower perceived cognitive load and more efficient gaze behavior. 
Crucially, predictive modeling is context-sensitive: the relationship between eye-tracking signals and cognitive states shifts across AI conditions. 
Finally, fusing eye-tracking features with user priors (demographics, AI literacy/experience, and propensity to trust technology) improves cross-participant generalization. These findings support condition-aware and personalized user modeling for cognitively aligned adaptive AI systems.
\end{abstract}

\makeatletter
\g@addto@macro{\@authornotes}{%
  \footnotetext{$^{*}$Corresponding authors: Abdallah El Ali; Saku Sugawara.}%
}
\makeatother

\maketitle


\vspace{-0.8mm}


\section{Introduction}

As AI becomes increasingly embedded in high-stakes human decision making~\cite{AI_assisted_decision_making}, understanding users’ moment-to-moment states, such as perceived cognitive load and decision confidence, is critical for building human-AI aligned interactive systems. 
In practice, the same user may respond differently~\cite{physio_search}, in attention allocation, confidence, and decisions, depending on \emph{how} AI is introduced and \emph{whether} the AI advice is correct~\cite{xsun_reasoning_chi26}. 
Prior work in human-AI interaction shows that perceived trustworthiness~\cite{Vereschak2024,confidence_effect_in_decision_1}, confidence in decision-making~\cite{confidence_effect_in_decision_1,confidence_rating}, and AI reliability and transparency~\cite{Wang2023-nm,explanation_ai_overreliance} influence how people engage with information, which can affect people's cognitive and attentional states.
These findings motivate user modeling approaches that sense and adapt to users' cognitive states during AI-assisted decision-making.


A promising signal source for such user modeling is eye tracking: gaze behavior and pupil-related measures have been used as behavioral correlates of cognitive and attentional effort in reading, reasoning and decision-making~\cite{eyetracking_survey,cognitive_pupil,physio_search,eye_movements_decision}. 
However, it remains unclear whether these signals provide a \emph{robust} basis for cognitive-state modeling when the context is influenced by AI conditions. 
Users may allocate cognitive attention differently, adopt different strategies, and experience different levels of confidence and effort depending on whether AI advice is present and whether it is correct or not (cf., hallucinations~\cite{Hallucin28:online,Hallucinations} or misinformation~\cite{yao2025factcheckingaigeneratednewsreports}). 
Most prior work analyzes gaze or physiological signals without explicitly accounting for the AI condition in cognitive state modeling~\cite{eyetracking_survey,sun_trust}. 
Moreover, AI-assisted decision-making exhibits substantial \emph{individual differences}~\cite{Vereschak2024}. 
User factors such as demographics, AI literacy and experience, and propensity to trust technology can influence reliance on AI and subjective judgments~\cite{Vereschak2024,explanation_ai_overreliance}. 
Yet prior work of eye-tracking-based user modeling rarely tests whether incorporating such user priors improves modeling in AI assistance, leaving open questions about cross-participant generalization and condition sensitivity of user modeling using eye-tracking signals.

To address these gaps, we investigate how eye-tracking signals relate to self-reported cognitive load, decision confidence, and decision performance under varying AI advice conditions, and whether user priors improve generalization to unseen users. 
We conduct a controlled lab study using factual verification tasks with $N=54$ participants under three within-subject conditions: (1) \textbf{No-AI}, (2) \textbf{Correct-AI} advice, and (3) \textbf{Incorrect-AI} advice.
For each factual claim, we created matched Correct and Incorrect AI advice, where Incorrect-AI \emph{flipped the recommendation relative to the ground-truth} while keeping length and format comparable to Correct-AI advice, and we verified intended correctness manipulation via internal validation.
For each trial, we collect self-reports and simultaneously record eye-tracking signals (i.e., gaze behaviors and pupil measures). 
We evaluate predictive models under leave-one-subject-out cross-validation (LOSOCV)~\cite{loso} to assess cross-participant generalization and report condition-stratified performance.


\begin{figure*}[!t]
\raggedright 
\includegraphics[width=0.992\textwidth]{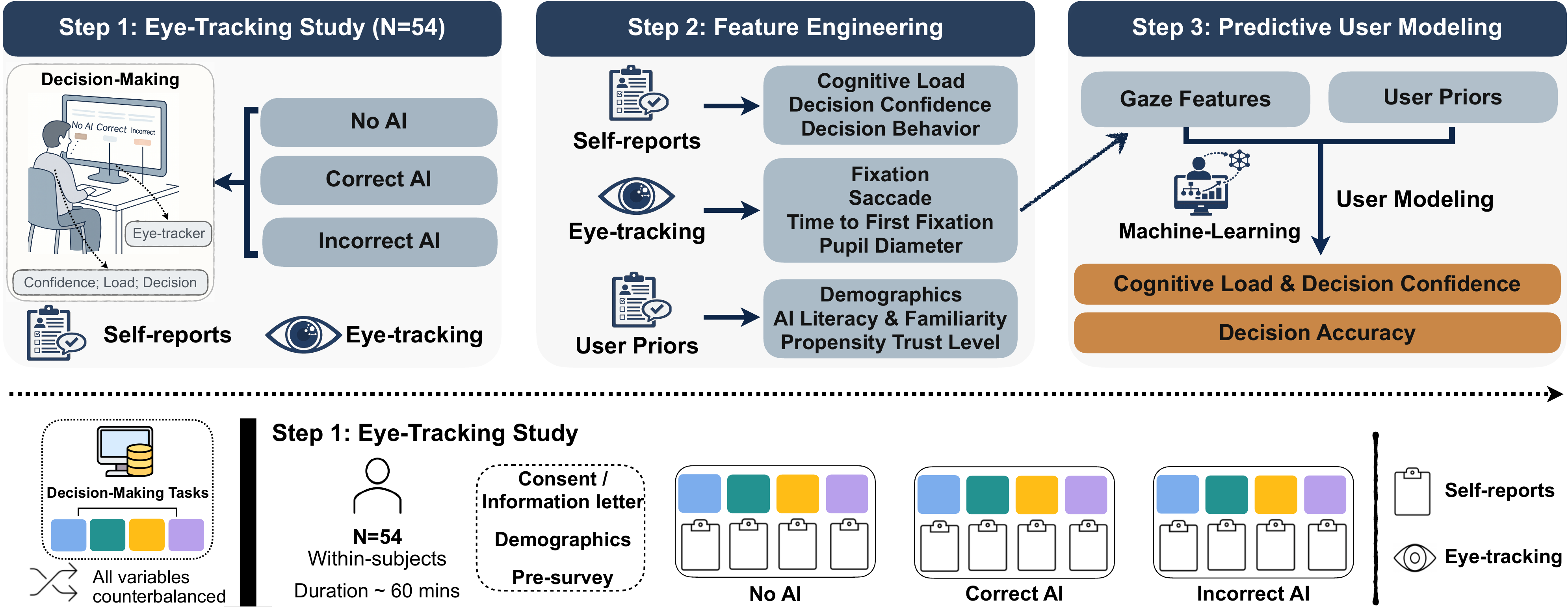}
\vspace{-1.6mm}
\caption{\textbf{Study overview and procedure.} 
\emph{(Top):} Three-step workflow. 
\emph{Step 1} collects eye-tracking signals and self-reports during factual verification under three within-subject AI conditions. 
\emph{Step 2} extracts trial-level eye-tracking signals and outcomes (cognitive load, decision confidence, and accuracy) together with participant-level user priors (e.g., demographics, AI literacy, propensity to trust). 
\emph{Step 3} trains machine learning models to predict users' cognitive load, decision confidence, and accuracy from eye-tracking features alone or fused with user priors, across AI conditions. 
\emph{(Bottom):} Study procedure in Step 1: consent and pre-survey, followed by counterbalanced trials spanning three AI conditions with concurrent eye-tracking and self-reports.}
\vspace{-1.4mm}
\label{fig:procedure}
\end{figure*}



Our investigation is guided by three research questions:
\begin{enumerate}[label=, leftmargin=-0.1mm, itemsep=0pt]
    \vspace{-0.82mm}
    \item \textit{\textbf{RQ1:} 
    How do AI presence and correctness affect users' cognitive load, decision confidence and accuracy, as well as eye-tracking patterns?
    }
    \vspace{1.0mm}
    \item \textit{\textbf{RQ2:} 
    Can models reliably predict self-reports (cognitive load and confidence) and decision accuracy from eye-tracking signals, and how does performance vary across AI conditions?
    }
    \vspace{1.0mm}
    \item \textit{\textbf{RQ3:} 
    Do user priors (demographics, AI literacy and experience, and propensity to trust technology) improve cross-participant generalization of user modeling when fused with eye-tracking features?
    }
    \vspace{-0.82mm}
\end{enumerate}



Our results reveal three key findings.
First, AI advice influences both explicit self-reports and implicit eye-tracking signals: compared to No-AI, AI advice increases decision confidence, and \textit{Correct-AI} is associated with lower perceived cognitive load and more efficient visual processing patterns.
Second, user modeling from eye tracking is \emph{condition-sensitive}: models trained on pooled data do not consistently generalize across conditions, and the best-performing features/models differ across conditions and predicted targets.
Third, fusing user priors (demographics, AI literacy and experience, and propensity to trust technology) with eye-tracking features improves cross-participant generalization.


This work makes three contributions:
(1) We characterize conditi-\break on-sensitive effects in self-reported cognitive states and eye-tracking patterns under varying AI assistance. 
(2) We systematically compare and evaluate pooled and condition-specific modeling strategies under AI assistance and correctness manipulation.  
(3) We show that fusing user priors with eye-tracking signals improves generalization, offering practical implications for cognitively-aligned user modeling and adaptive AI-assisted decision-making systems that account for both AI conditions and individual differences.

\section{Related Work}

\subsection{AI-assisted decision-making}


AI is increasingly used to support human decision-making, and carefully designed assistance can improve decision performance and trust in some settings~\cite{confidence_effect_in_decision_1,confidence_effect_in_decision_2}. 
At the same time, AI support can induce reliance and errors: automation bias and related overreliance effects lead users to accept incorrect advice or miss AI failures, especially when reliability cues are unclear~\cite{bias_ai,explanation_ai_overreliance}. 
A growing body of HAI work shows that users’ judgments and reliance are affected by how AI \emph{communicates} its trustworthiness, e.g., certainty indicators, transparency, and explanations, and that these cues can have context-dependent effects on perceived trust and downstream decisions~\cite{confidence_effect_in_decision_1,Wang2023-nm,Springer2020,responsible_ai}. 
Recent studies further highlight that AI-provided confidence can shift users’ \emph{self-confidence} and its calibration, potentially persisting beyond the interaction~\cite{measure_confidence}, and that explanation quality or even misleading/imperfect explanations can meaningfully alter users’ acceptance and trust~\cite{10.1145/3290605.3300717,10.1145/3375627.3375833,incorrect_xai_1}.

\vspace{-1.7mm}
\subsection{Gaze signal for cognitive state inference}

Eye-tracking provides a non-invasive window into cognitive processing and decision-making~\cite{eye_movements_decision}. 
A large body of psychology and HCI research has established that gaze behaviors such as longer fixations, and physiological signals such as pupil dilation can reflect processing difficulty, mental effort, and cognitive load~\cite{eyetracking_survey,eyetracking_cognitive_2,physio_search}. 
Besides, prior work has operationalized eye-tracking features (e.g., fixation/saccade statistics, pupil dynamics) as inputs for machine learning models to infer users’ perceived trust~\cite{Ahmad_Muneeb}, sources~\cite{SUMER2021106909,Abdrabou_Yasmeen,sun_trust}, cognitive load, and biases~\cite{Parikh2018EyeGF,eyetracking_cognitive_2,eyetracking_cognitive_1,physio_search} in response to AI-generated content.
These results suggest that gaze signals can support low-burden sensing for user modeling when the mapping between attention/physiology and subjective state is relatively stable.

However, applying gaze-based inference to AI-assisted decision-making raises a key challenge of \emph{signal attribution}. 
In such tasks, an observed change in gaze or pupil dilation may reflect multiple overlapping sources: inherent evidence difficulty, additional effort spent interpreting AI advice or transparency cues, or cognitive conflict when AI suggestions are misleading~\cite{eyetracking_cognitive_1}. 
Meanwhile, AI assistance can also shift users’ decision strategies and attention allocation as a function of perceived reliability, trustworthiness and reliance~\cite{confidence_effect_in_decision_1,Vereschak2024,explanation_ai_overreliance}. 
This implies that the mapping from implicit gaze signals to subjective states may \emph{change with AI context} rather than remain fixed, making it unclear whether models trained in one setting will generalize across different reliability conditions or across users. 
Our work directly addresses this gap by testing gaze-based inference of cognitive load and decision confidence under controlled No-AI, Correct-AI, and Incorrect-AI contexts.


\subsection{Personalized and multimodal user modeling}


UMAP research emphasizes modeling users' states and individual differences to support personalization and adaptive interaction. 
Personalized user modeling aims to capture individual differences (e.g., skills, attitudes, trust) so that systems can adapt to \emph{who} the user is, not only \emph{what} they do in a single interaction. 
In AI-assisted decision-making, such personalization is especially important because users vary in AI literacy and experience~\cite{ai_literacy} and in their propensity to trust automation/technology~\cite{ppt,measure_trust_automation}, which can shape reliance, confidence, and susceptibility to overreliance. 
Recent HAI work further argues that responsible AI experiences should account for heterogeneity in users’ beliefs and interpretations of system cues~\cite{responsible_ai,ux_responsible}.

A complementary line of work uses data to infer user states by combining behavioral and physiological signals and learning predictive models~\cite{Ahmad_Muneeb,Akash_Kumar,Ajenaghughrure_modeling,eyetracking_cognitive_2,sun_trust,physio_search}. 
However, sensing-based models emphasize within-subjects performance, while deployable settings require generalization to unseen users (a cold-start problem). 
This motivates us to fuse \emph{dynamic interaction signals} (e.g., gaze) with \emph{stable user priors} (e.g., demographics, AI literacy, trust propensity) to improve cross-participant robustness and calibration. 


\section{Methods}

We conducted experiments with the following three steps (Fig.~\ref{fig:procedure}).


\vspace{-1.4mm}
\subsection{Empirical study with eye-tracking}


\begin{table}[!b]
\centering
\footnotesize
\renewcommand{\arraystretch}{0.9}
\vspace{-4.0mm}
\begin{tabularx}{\columnwidth}{p{2.6cm} >{\raggedright\arraybackslash}p{2.5cm} >{\raggedleft\arraybackslash}p{2.0cm}}
\toprule
\textbf{Demog.} & \textbf{Categ.} & \textbf{Participants (\%)} \\
\midrule
Gender &  &   \\
\hline
 & Female & 48 (88.9\%) 
\\ 
 & Male & 6 (11.1\%) 
\\
\hline
Age & & 
\\
 & 18-24 & 36 (66.7\%) 
\\
 & 25-34 & 16 (29.6\%) 
\\
 & 35-44 & 2 (3.7\%) 
\\
\hline
Education & & 
\\
 & High school & 8 (14.8\%) 
\\
 & Bachelor’s degree & 28 (51.9\%) 
\\
 & Master’s degree & 17 (31.5\%) 
\\
 & Doctorate & 1 (1.9\%) 
\\ \hline
AI Familiarity & & 
\\
 & Never & 0 (0.0\%) 
\\
 & Slightly & 5 (9.3\%) 
\\
 & Moderately & 20 (37.0\%) 
\\
 & Very & 21 (38.9\%) 
\\
 & Extremely & 8 (14.8\%) 
\\
\bottomrule
\end{tabularx}
\vspace{-0.0mm}
\caption{Characteristics of the participants in Study of Step 1.}
\vspace{-7.6mm}
\label{table:lab_demographics}
\end{table}

\subsubsection{Study design and conditions}

As shown in Fig.~\ref{fig:procedure}, we used a within-subjects design in which each participant completed 12 factual verification trials as the decision-making. 
In each trial, participants read a claim and its supporting evidence (adapted from dataset StrategyQA~\cite{strategyqa}) and then made a binary \textit{True/False} judgment. 
To manipulate the decision context, each trial was assigned to one of three AI assistance conditions: 
(1) \textit{No-AI}: no AI advice was shown; 
(2) \textit{Correct-AI}: correct AI-generated advice shown; 
(3) \textit{Incorrect-AI}: an incorrect AI suggestion was shown. 
Trials were evenly distributed across three AI conditions (No-AI / Correct-AI / Incorrect-AI; 4 each).
Each participant experienced all conditions in a counterbalanced order to mitigate sequence effects.


\subsubsection{Participants}
A priori G*Power~\cite{gpower} analysis suggested 32 participants to detect a medium effect (Cohen’s $d=0.25$) with 80\% power. 
We recruited \textit{54} participants via the institute's subject pool. 
All participants received compensation for a 60-minute in-lab session.
The study was approved by the institute's ethical committee.
Participant demographics are reported in Table~\ref{table:lab_demographics}.



\subsubsection{Materials: stimuli and apparatus}

We developed a custom web-based interface to present stimuli and collect participants’ responses. 
Each task consisted of a binary fact-checking claim (True/False) with supporting context. 
The interface displayed the LLM-generated answer. 
After each trial, participants provided their self-reported decisions and ratings.
The factual claims were adapted from the public StrategyQA dataset~\citep{strategyqa}.
For each claim, we prepared two AI advice variants matched in length and format: (i) Correct-AI advice aligned with the ground-truth, and (ii) Incorrect-AI advice whose recommendation was intentionally flipped to be incorrect while keeping surface-level style comparable. 
We conducted an internal check to ensure the manipulation of advice.
Examples of the claim and AI advice are added in the supplementary material.

Fig~\ref{fig:lab_gaze_heatmap} shows that experiment was run on a Windows PC with a 27" monitor (1920 x 1080, 100 Hz). A Tobii Pro Fusion eye tracker (60\,Hz) recorded eye-tracking signals~\citep{tobii_pro}. 
Stimuli and questionnaires were presented through PsychoPy~\citep{psychopy}, which synchronized experimental tasks with recording to ensure precise alignment.


\begin{figure}[!t]
\raggedright 
\includegraphics[width=0.480\textwidth]{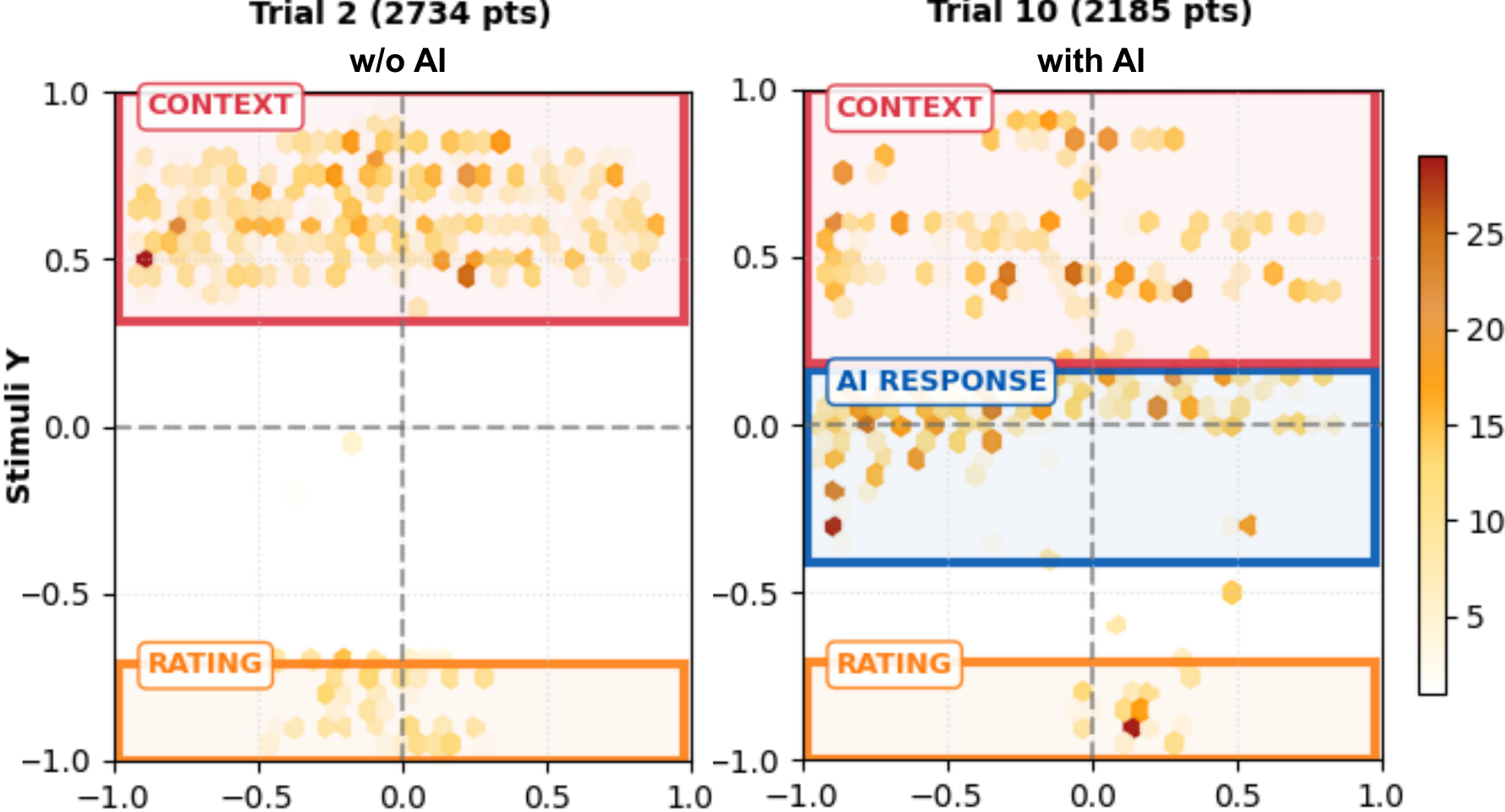}
\vspace{-3.6mm}
\caption{
Heatmaps of gaze per trial from conditions with or without AI. 
Higher density indicates greater visual attention.}
\label{fig:lab_gaze_heatmap}
\vspace{-3.6mm}
\end{figure}


\subsubsection{Measure}
\label{measures}

\textbf{User priors.} 
\label{user_priors_measures}
Prior to the study, participants completed a short questionnaire capturing demographics and user priors related to human-AI interaction (Table~\ref{table:lab_demographics}), including prior experience with AI, AI literacy~\cite{ai_literacy} and propensity to trust technology (PPT)~\citep{ppt} (used as participant-level features in modeling).

\textbf{Self-reports.} 
\label{self_reports_measures}
After each task, participants provided the following responses:
(1) \textbf{Decision:} Participants made a binary decision on the given factual claim (True or False). 
(2) \textbf{Decision confidence:} Participants rated how confident they were in their decision on a 7-point Likert scale (from ``Not confident at all'' to ``extremely confident'')~\citep{confidence_rating}.
(3) \textbf{Cognitive load:} Participants reported perceived cognitive load after each trial using 7-point Likert scales (1=low, 7=high) adapted from~\cite{cognitive_load_rating} (Cronbach’s alpha=.79). 
An additional \textbf{Manipulation check} was used after each AI trial to check whether participants perceived the AI response as correct or not.

\textbf{Eye-tracking signals.} 
\label{gaze_measures}
We recorded gaze behaviors and physiological pupil measures throughout trials. 
We derived standard eye-tracking features, including: \emph{fixation count and duration~\cite{fixations}, saccade count and length~\cite{saccades}, time to first fixation (TTFF), and pupil diameter~\cite{eyetracking_cognitive_2,physio_search}}.
Areas of interest (AOIs) were defined as Fig.~\ref{fig:lab_gaze_heatmap} (i.e., claim and evidence text as AOI-Context, AI advice panel when present as AOI-Advice, and the rating panel as AOI-Rating).


\subsubsection{Procedure}
As shown in Fig.~\ref{fig:procedure}, participants provided consent and completed a short pre-survey. 
They then sat in front of an eye-tracking setup and completed a standard eye-tracker calibration routine~\cite{eyetracking_methods}. 
The study began with a brief tutorial trial to familiarize participants with the study.
Each participant completed 12 trials. 
In each trial, participants read claim and supporting evidence, viewed AI advice when present, and made a binary decision. 
They then reported their decision confidence and perceived cognitive load. 


\subsubsection{Data analysis}
\label{data_analysis}
We analyzed the data to address RQ1.

\textbf{Preprocessing of eye-tracking data.} 
We applied standard preprocessing~\cite{eyetracking_methods}. 
Gaze patterns and pupil diameter were calculated by the I-VT signal filter~\cite{tobii_filter}. We excluded trials that did not meet predefined quality criteria (gaze validity $<$ 80\%). 

\textbf{Condition-aware effects.} 
We analyzed self-reports (cognitive load, confidence), decision accuracy by repeated measures ANOVA analysis~\cite{anova} with FDR-correction~\citep{fdr_bh} across AI conditions to answer RQ1.
Eye-tracking data was analyzed by MixedLM~\citep{mixedLM} across three AOIs. 
Both self-reports and eye-tracking were further analyzed by post-hoc pairwise t-test~\cite{t-test}. 




\subsection{Feature engineering for predictive modeling}
\label{feature_engineering}

For each trial $i$, we extracted trial-level gaze and pupil features (see Sec.~\ref{gaze_measures}). 
Each trial is represented by an eye-tracking feature vector $\mathbf{g}_i \in \mathbb{R}^{d}$ with an associated AI condition label $c_i \in \{\textit{No-AI}, \textit{Correct-AI}, \textit{Incorrect-AI}\}$. 
Specifically, we computed standard eye-tracking metrics per AOI, including fixation count and duration, saccade count and length, time-to-first-fixation (TTFF), and physiological feature pupil diameter per AOI.  
Trial-level outcomes include participants' decision accuracy $\text{accuracy}_i$ and self-reported confidence $\text{confidence}_i$ and cognitive load $\text{load}_i$. 
In addition, each participant $u$ provides user-level priors $\mathbf{p}_u$ (see Sec.~\ref{user_priors_measures}), which are constant across that participant's trials. 
We use these priors either alone or fused with eye-tracking features for cross-participant modeling.
Thus, our dataset consists of trial-level features:
\[
\mathcal{D}=\left\{\left(\mathbf{c}_i,\ \mathbf{g}_i,\ p_{u},\ \text{load}_i,\ \text{confidence}_i,\ \text{accuracy}_i\right)\right\}_{i=1}^{N},
\]


\subsection{Conditional and personalized user modeling}
\label{modeling}

We formulate user modeling as a supervised prediction task: given eye-tracking features and user priors, we predict participants' self-reported \textit{cognitive load}, \textit{decision confidence} and decision accuracy.

\textbf{Modeling across AI conditions (RQ2).}
Each trial is labeled with an AI condition $c_i \in \{\textit{No-AI}, \textit{Correct-AI}, \textit{Incorrect-AI}\}$.
We evaluate three modeling strategies:
\textbf{(1) Pooled (All):} one model trained on all trials across conditions.
\textbf{(2) Condition-specific:} three models trained within each condition.
\textbf{(3) Condition-aware mixture-of-experts (MoE):} three condition-specific experts trained separately, where inference routes each test trial to the expert corresponding to its known condition $c_i$.
\textit{To address \textbf{RQ2}}, we trained machine learning (ML) models to learn the mapping:
\[
f: \mathbf{g}_i \mapsto \left\{\left(\text{load}_i,\text{confidence}_i,\text{accuracy}_i\right)\right\}_{i=1}^{N},
\]

\textbf{Personalized modeling with priors (RQ3).}
We examined ML prediction with the value of user priors across AI conditions.
We evaluate three feature settings: 
\textbf{(i) Eye-tracking-only}: $\mathbf{g}_i$, include both gaze behaviors (i.e., fixation counts/duration, saccade counts/length, TTFF) and physiological pupil diameter;
\textbf{(ii) User Priors-only}: $\mathbf{p}_{u_i}$, include demographics, AI literacy and experience, and propensity to trust technology (Sec.~\ref{measures}).;
\textbf{(iii) Multimodal Fusion}: $[\mathbf{g}_i + \mathbf{p}_{u_i}]$.
This ablation isolates how much predictive signal comes from eye-tracking versus user priors, and whether they provide complementary information.
Thus, we trained machine learning (ML) models to learn the mapping:
\[
f: (\mathbf{g}_i, \mathbf{p}_{u_i}) \mapsto \left\{\left(\text{load}_i,\text{confidence}_i,\text{accuracy}_i\right)\right\}_{i=1}^{N},
\]

\textbf{Setup and evaluation.} 
We binarize the self-reported \textit{cognitive load} and \textit{decision confidence} for classification. For each training fold, we compute the median of the predicted target on the \emph{training participants only} and use it as the threshold: trials with values above (or equal to) the training median are labeled as \textit{high}, and those below as \textit{low}, to ensure label balance.
We report Accuracy, F1~\cite{f1}, and AUC~\cite{auc} as the evaluation metrics. 



\section{Results}




\subsection{Effects of AI-assisted decision-making - \textit{RQ1}}



\subsubsection{Self-reports} 


\newlength{\colA}\setlength{\colA}{0.14\columnwidth} 
\newlength{\colB}\setlength{\colB}{0.14\columnwidth} 
\newlength{\colC}\setlength{\colC}{0.38\columnwidth} 
\newlength{\colD}\setlength{\colD}{0.07\columnwidth} 
\newlength{\colE}\setlength{\colE}{0.08\columnwidth} 

\begin{table}[htb]
\centering
\small
\renewcommand{\arraystretch}{0.88} 
\begin{tabular*}{\columnwidth}{@{\extracolsep{\fill}}p{\colA}p{\colB}p{\colC}p{\colD}p{\colE}@{}}
\toprule
\textbf{Measure} & \textbf{ANOVA} & \textbf{Pairwise (mean)} & \textbf{$p$} & \textbf{Effect} \\
\midrule
\multirow{3}{*}{Load}
  & \multirow{3}{*}{.001 (.33)}
  & No vs. Corr (3.56 vs. 3.18)       & \sigtwo{.010} & .35 \\
  &  & No vs. Incorr (3.56 vs. 3.38)     & .200 & .18 \\
  &  & Corr vs. Incorr (3.18 vs. 3.38)  & \sigtwo{.040} & .28 \\
\midrule
\multirow{3}{*}{Confidence}
  & \multirow{3}{*}{<.000 (.42)}
  & No vs. Corr (5.22 vs. 5.93)    & \sigtwo{<.001} & .60 \\
  &  & No vs. Incorr (5.22 vs. 5.79)    & \sigtwo{.002} & .45 \\
  &  & Corr vs. Incorr (5.93 vs. 5.79)  & .220 & .17 \\
\midrule
\multirow{3}{*}{Accuracy}
  & \multirow{3}{*}{.500 (.12)}
  & No vs. Corr (.78 vs. .77)        & .830 & .03 \\
  &  & No vs. Incorr (.78 vs. .82)     & .270 & .15 \\
  &  & Corr vs. Incorr (.77 vs. .82) & .080 & .24 \\
\bottomrule
\end{tabular*}
\caption{Results from ANOVA analysis (p value with effect size) and pairwise comparisons with mean values. Measures are cognitive load, decision confidence and decision accuracy.}
\label{tab:anova_pairwise}
\vspace{-6.2mm}
\end{table}


Table~\ref{tab:anova_pairwise} reports the ANOVA and pairwise comparisons across No-AI, Correct-AI, and Incorrect-AI conditions. 
The manipulation check was significant ($p<.001$), indicating that participants perceived the intended advice correctness differences.  

\textbf{Cognitive load}
differed significantly by condition (ANOVA: $p=.001$, $\eta^2=.33$). 
Pairwise tests showed lower load in \textit{Correct-AI} than \textit{No-AI} (3.18 vs.\ 3.56, $p$=.010, effect=$.35$) and than \textit{Incorrect-AI} (3.18 vs.\ 3.38, $p$=.040, effect=$.28$). 
The difference between \textit{No-AI} and \textit{Incorrect-AI} was not significant (3.56 vs.\ 3.38, $p$=.200).

\textbf{Decision confidence}
differed significantly by condition (ANOVA: $p<.001$, effect=$.42$). 
Confidence was higher with AI advice than without AI for both Correct-AI ($p$<.001, effect=$.60$) and Incorrect-AI ($p$=.002, effect=$.45$), with no significant difference between \textit{Correct-AI} and \textit{Incorrect-AI} conditions ($p$=.220).
\textbf{Decision accuracy} also did not differ significantly by condition (ANOVA: $p$=.500, effect=.12).


\begin{figure}[!ht]
    \centering
    \includegraphics[width=0.98\linewidth]{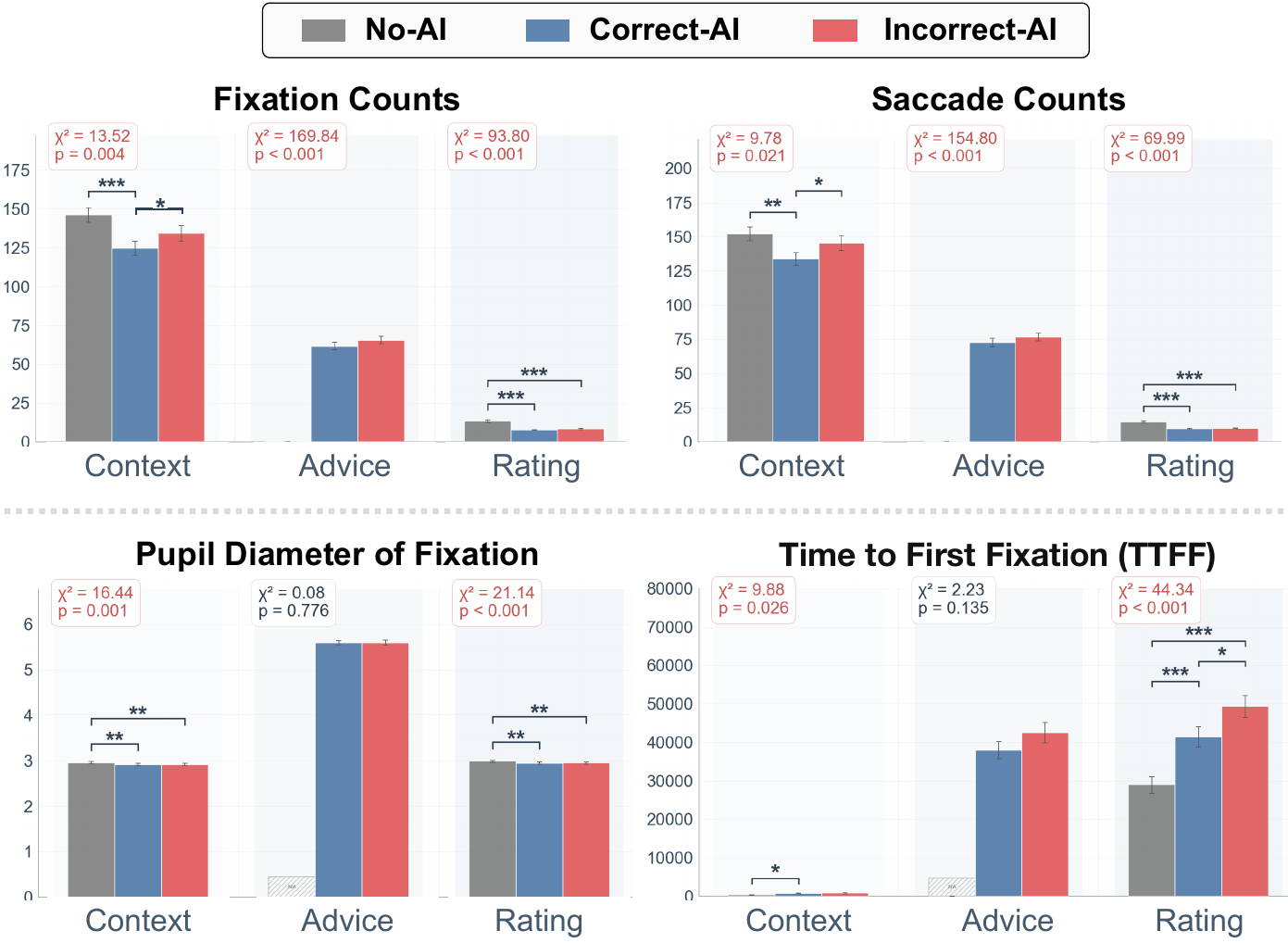}
    \vspace{-3.2mm}
    \caption{
    Gaze features (i.e., Fixation/Saccade Count, Pupil Diameter, and Time-to-First-Fixation across AOIs (Context, AI Advice, and User Rating areas). 
    Bars show the mean. Values in upper-left ($X^2$, p) are results from MixedLM model (FDR-corrected), and brackets indicate significant pairwise t-tests $(*p < .05, **p < .01, ***p < .001)$.
    }
    \vspace{-4.6mm}
    \label{fig:lab_gaze_analysis}
\end{figure}


\subsubsection{Gaze patterns vary across conditions} 


We processed the raw eye-tracking signal~\citep{eyetracking_methods} and analyzed it to complement self-reported outcomes and test how presence and correctness of AI advice shape visual attention (Fig.~\ref{fig:lab_gaze_analysis}) across conditions and AOIs (see Sec.~\ref{data_analysis}).

\textbf{No-AI.}  
Participants showed more and longer fixations and saccades, as well as larger pupil diameter on the context (AOI-Context) and rating items (AOI-Rating), than in conditions with AI. 
Participants also oriented most quickly to rating items (shorter time to first fixation). 
This pattern suggests that participants relied primarily on the context for decisions with higher uncertainty, consistent with the lowest decision confidence and highest cognitive effort.

\textbf{Correct-AI.}  
With correct AI advice, participants made significantly fewer fixations and saccades on the evidence (AOI-Context) and rating items (AOI-Rating) than in No-AI condition. 
Compared to Incorrect-AI, they showed lower saccade activity in advice region (AOI-Advice), suggesting reduced effort when processing advice. These patterns are consistent with self-reports showing higher decision confidence and lower cognitive load in Correct-AI condition.

\textbf{Incorrect-AI.}  
Participants showed more visual processing on AOI-Context than Correct-AI condition, reflected by higher fixations and saccades. 
Participants also took longer to shift attention to the AOI-Rating, with the \emph{longest} TTFF in Incorrect-AI. 
It indicates that Incorrect-AI is associated with increased cognitive effort to evidence when processing incorrect AI advice. 


\vspace{-1.2mm}
\subsection{User modeling via eye-tracking alone - \textit{RQ2}}

\begin{table*}[!ht]
\centering
\footnotesize
\renewcommand{\arraystretch}{0.74}
\setlength{\tabcolsep}{2pt}

\newcolumntype{L}[1]{>{\raggedright\arraybackslash}p{#1}}
\newcolumntype{Y}{>{\raggedright\arraybackslash}X}

\begin{tabularx}{\textwidth}{@{} L{1.4cm} L{0.86cm} *{11}{Y} @{}}
\toprule
\multirow{2}{*}{\textbf{Targets}}
& \multirow{2}{*}{\textbf{Models}}
& \multicolumn{5}{c}{\colorbox{gray!18}{\textbf{Eye-tracking Only}}}
& \multicolumn{1}{c}{\colorbox{blue!10}{\textbf{User Priors}}}
& \multicolumn{5}{c}{\colorbox{red!10}{\textbf{Multimodal Fusion}}}\\
\cmidrule(lr){3-7} \cmidrule(lr){8-8} \cmidrule(lr){9-13}
& &
\textbf{All} & \textbf{No} & \textbf{Corr} & \textbf{Incorr} & \textbf{MoE}
& \textbf{All}
& \textbf{All} & \textbf{No} & \textbf{Corr} & \textbf{Incorr} & \textbf{MoE} \\
\midrule

\multirow{7}{*}{\textbf{\makecell[l]{Cognitive\\Load}}}
& LR  & .61/.65/.64 & .61/.57/.68 & .50/.51/.52 & \underline{.63/.60/.62} & .58/.63/.63
      & .65/.69/.68
      & .66/.71/.73 & .65/.64/.70 & .58/.63/.62 & \underline{.70/.72/.74} & .65/.71/.69 \\
& SVM & .62/.76/.51 & .57/.64/.52 & .60/.75/.55 & .49/.45/.48 & \underline{.62/.76/.52}
      & .68/.75/.63
      & \underline{.66/.76/.51} & .57/.63/.52 & .60/.75/.56 & .49/.45/.48 & .61/.76/.52 \\
& RF  & \underline{.66/.74/.68} & .64/.64/.68 & .58/.67/.58 & .60/.60/.61 & .64/.74/.62
      & .65/.71/.68
      & \underline{.71/.78/.76} & .67/.67/.72 & .59/.69/.62 & .63/.62/.66 & .65/.75/.67 \\
& ET  & \underline{.65/.74/.68} & .63/.62/.68 & .58/.68/.57 & .55/.51/.61 & .62/.72/.63
      & .68/.74/.68
      & .65/.81/.79 & .\underline{69/.69/.74} & .63/.72/.67 & .68/.75/.72 & .68/.75/.72 \\
& AdaB& \underline{.62/.73/.63} & .63/.62/.67 & .60/.75/.52 & .57/.51/.57 & .61/.70/.60
      & .66/.73/.67
      & .65/.74/.67 & \underline{.67/.68/.68} & .60/.72/.52 & .57/.52/.56 & .63/.71/.63 \\
& XGB & \best{.66/.75/.69} &.61/.64/.66 & .57/.69/.60 & .62/.59/.63 & .60/.70/.60 
      & \best{.68/.73/.67}
      & \underline{.69/.77/.74} & .65/.66/.70 & .60/.71/.60 & .56/.56/.62 & .63/.71/.62 \\
& MLP & \underline{.64/.74/.62} & .58/.60/.62 & .54/.65/.52 & .57/.60/.60 & .57/.66/.59
      & .64/.74/.67
      & \best{.73/.79/.76} & .61/.59/.67 & .46/.51/.47 & .62/.61/.65 & .62/.73/.62 \\
\midrule

\multirow{7}{*}{\textbf{\makecell[l]{Decision\\Confidence}}}
& LR  & .62/.69/.63 & \underline{.63/.66/.68} & .60/.68/.58 & .59/.68/.59 & .58/.66/.57
      & .57/.64/.61
      & .62/.68/.66 & \underline{.69/.72/.73} & .58/.68/.54 & .62/.71/.60 & .60/.68/.60 \\
& SVM & .65/.78/.59 & .66/.70/.56 & .67/.79/.59 & \underline{.68/.81/.54} & .65/.77/.52
      & .65/.78/.57
      & .63/.76/.60 & .66/.70/.56 & \underline{.68/.80/.60} & .68/.81/.54 & .65/.77/.52 \\
& RF  & .65/.76/.67 & .67/.67/.70 & \underline{.71/.81/.67} & .67/.78/.65 & .67/.77/.66
      & .59/.66/.63
      & .66/.77/.68 & .67/.71/.73 & \underline{.71/.81/.69} & .67/.78/.65 & .67/.77/.67 \\
& ET  & .65/.77/.66 & .63/.70/.69 & \underline{.71/.81/.66} & .66/.77/.65 & .66/.77/.65
      & .65/.75/.63
      & .67/.77/.66 & .68/.71/.74 & \underline{.74/.83/.70} & .68/.77/.61 & .67/.77/.68 \\
& AdaB& .67/.79/.64 & .59/.66/.61 & \underline{.70/.82/.65} & .67/.80/.52 & .61/.72/.59
      & .64/.78/.59
      & .68/.78/.65 & .62/.68/.65 & \underline{.70/.81/.64} & .66/.79/.53 & .63/.74/.61 \\
& XGB & .63/.75/.65 & .61/.67/.69 & \best{.71/.81/.69} & .67/.79/.62 & .65/.75/.64
      & \best{.65/.78/.62}
      & .64/.75/.68 & .68/.72/.71 & \best{.74/.83/.72} & .65/.77/.60 & .63/.73/.62 \\
& MLP & .65/.76/.64 & .64/.70/.65 & \underline{.65/.77/.54} & .63/.74/.57 & .59/.69/.58
      & .62/.75/.54
      & .65/.75/.63 & .66/.70/.71 & \underline{.74/.83/.62} & .66/.78/.58 & .66/.76/.64 \\
\midrule

\multirow{7}{*}{\textbf{\makecell[l]{Decision\\Accuracy}}}
& LR  & .57/.69/.59 & .56/.67/.53 & .52/.65/.51 & .54/.68/.48 & \underline{.61/.73/.54}
      & .60/.71/.62
      & .64/.75/.62 & .59/.71/.54 & .66/.77/.57 & \underline{.68/.79/.55}& .67/.78/.60 \\
& SVM & .79/.88/.49 & .78/.87/.41 & .78/.87/.44 & \underline{.83/.91/.46} & .79/.88/.47
      & .80/.88/.61
      & .79/.88/.49 & .78/.87/.41 & .77/.87/.41 & \underline{.83/.91/.49} & .79/.88/.47 \\
& RF  & .79/.88/.62 & .76/.86/.47 & .74/.85/.55 & \underline{.82/.90/.57} & .79/.88/.58
      & .67/.77/.64
      & .79/.88/.64 & .76/.86/.51 & .76/.86/.56 & \underline{.84/.91/.57} & .79/.88/.61 \\
& ET  & .79/.88/.62 & .76/.86/.61 & .75/.86/.50 & \best{.82/.90/.61} & .79/.88/.60
      & .80/.88/.62
      & .79/.77/.66 & .78/.87/.61 & .74/.84/.60 & \underline{.81/.89/.57} & .77/.87/.61 \\
& AdaB& .79/.88/.57 & .74/.85/.54 & .76/.87/.50 & \underline{.82/.90/.53} & .75/.85/.55 & .79/.88/.60 & .79/.88/.56 & .76/.86/.51 & .76/.87/.52 & \underline{.83/.90/.53} & .73/.84/.55 \\
& XGB & .78/.88/.62 & .77/.87/.58 & .73/.84/.47 & .82/.90/.59 & .74/.85/.57
      & \best{.80/.88/.63}
      & .78/.88/.65 & .76/.86/.57 & .75/.86/.47 & .83/.91/.57 & .75/.85/.56 \\
& MLP & \underline{.79/.88/.54} & .72/.83/.51 & .67/.79/.43 & .79/.88/.48 & .70/.82/.53
      & .79/.88/.58
      & .79/.88/.55 & .73/.84/.53 & .71/.82/.54 & \best{.85/.91/.58} & .78/.87/.62 \\
\bottomrule
\end{tabularx}

\vspace{-0.0mm}
\caption{
\textbf{Predictive modeling results under different feature sets and AI conditions.} 
Binary classification performance \textcolor{red}{(Accuracy/F1/AUC)} for predicting \textit{cognitive load}, \textit{decision confidence and accuracy}. 
We compare three feature settings: \textit{Eye-tracking only}, \textit{User priors only}, and \textit{Multimodal fusion} (eye tracking + user priors). 
For each predicted target, we report overall performance on pooled trials (\textit{All}) and condition-stratified performance on specific \textit{No-AI}, \textit{Correct-AI}, and \textit{Incorrect-AI} trials; \textit{MoE} denotes a condition-aware mixture-of-experts model. 
Boldface indicates the best condition (All/No-AI/Correct-AI/Incorrect-AI/MoE) for each model within a given feature setting.
Red highlights the best-performing values for each predicted target within that given feature setting across conditions.
Model acronyms: LR (Logistic Regression); SVM (Support Vector Machine); RF (Random Forest); ET (ExtraTrees); AdaB (AdaBoost); XGB (XGBoost); MLP (Multi-Layer Perceptron).
}
\label{tab:res_predictions}
\vspace{-4.6mm}
\end{table*}

\begin{table}[t]
\centering
\footnotesize
\renewcommand{\arraystretch}{0.74}
\setlength{\tabcolsep}{2pt}
\newcolumntype{L}[1]{>{\raggedright\arraybackslash}p{#1}}
\newcolumntype{Y}{>{\raggedleft\arraybackslash}X}

\begin{tabularx}{\columnwidth}{@{} L{1.20cm} L{2.28cm} L{1.50cm} L{1.70cm} Y @{}}
\toprule
\textbf{Model} & \textbf{Predicted Target} & \textbf{Best Feature} & \textbf{Best Condition} & \textbf{Acc/F1/AUC} \\
\midrule
\multirow{3}{*}{LogReg}
  & Cognitive Load      & Fusion & Incorrect-AI & .70 / .72 / .74 \\
  & Decision Confidence & Fusion & No-AI        & .69 / .72 / .73 \\
  & Decision Accuracy   & Fusion & Incorrect-AI & .67 / .79 / .55 \\
\midrule
\multirow{3}{*}{ExtraTrees}
  & Cognitive Load      & Fusion & Pooled (All)   & .69 / .69 / .74 \\
  & Decision Confidence & Fusion & Correct-AI     & .74 / .83 / .70 \\
  & Decision Accuracy   & Fusion & Incorrect-AI   & .82 / .90 / .61 \\
\midrule
\multirow{3}{*}{XGBoost}
  & Cognitive Load      & Fusion & Pooled (All)   & .73 / .79 / .73 \\
  & Decision Confidence & Fusion & Correct-AI     & .74 / .83 / .72 \\
  & Decision Accuracy   & Fusion & Incorrect-AI   & .83 / .91 / .57 \\
\midrule
\multirow{3}{*}{MLP}
  & Cognitive Load      & Fusion & Pooled (All)   & .73 / .79 / .76 \\
  & Decision Confidence & Fusion & Correct-AI     & .74 / .83 / .62 \\
  & Decision Accuracy   & Fusion & Incorrect-AI   & .85 / .91 / .58 \\
\bottomrule
\end{tabularx}
\caption{\textbf{Summary of best-performing settings.} 
We report the feature set and AI condition that yields the best performance. 
}
\label{tab:res_predictions_best}
\vspace{-8.6mm}
\end{table}


Table~\ref{tab:res_predictions} (Eye-tracking Only part) answers \textbf{RQ2} on whether eye-tracking signals alone can reliably predict (i) participants’ self-reported \textit{cognitive load} and \textit{decision confidence}, (ii) \textit{decision accuracy}, and how predictions vary across AI conditions.
We compare three modeling settings (Sec.\ref{modeling}): (1) \textit{pooled} training over all AI conditions (\textit{All}), (2) \textit{condition-specific} training within each AI condition (No-AI, Correct-AI, Incorrect-AI), and (3) a condition-aware \textit{MoE}.

Across predictive targets, eye-tracking signals are \emph{most predictive of decision accuracy}.
Under pooled training (\textit{All}), all models (except LR) reach high accuracy ($\sim0.79$), indicating that eye-tracking carries robust information about whether a decision is correct.
However, the best performance often emerges under \emph{condition-specific} training rather than pooling: for some models, accuracy increases to low $\sim0.8$ range (e.g., $\sim0.82$ (ET) - $\sim0.83$ (SVM)) in certain AI conditions.
This pattern suggests that performance of modeling from eye-tracking is not fully invariant across conditions.

Predicting self-reported \textbf{cognitive load} and \textbf{confidence} from eye-tracking alone is more challenging.
Overall accuracies for these predictions are lower and show larger fluctuations across AI conditions than predicting decision accuracy.
For cognitive load, models achieve moderate performance (roughly $\sim0.66$ accuracy), and the pooled training settings achieved the best performance across most models, implying that cognitive load inferred from eye-tracking signals is sensitive to contextual factors.
For decision confidence, predictive performance varies more noticeably by condition: most models achieve their highest confidence prediction in Correct-AI condition, while performance can drop in No-AI or Incorrect-AI settings for several models, consistent with the idea that misleading AI advice can disrupt the stability of gaze-based confidence cues.

These results answer \textbf{RQ2} in two ways.
First, eye-tracking signals alone can predict \textit{decision accuracy} relatively reliably, whereas self-reports are moderately predictable.
Second, performance varies across AI conditions, and pooled (\textit{All}) training does not always dominate condition-specific modeling.
This provides direct evidence of \emph{conditional heterogeneity}: user modeling by eye-tracking cues in AI-assisted decision-making should explicitly account for AI conditions (e.g., whether AI advice is present and whether it is correct), rather than assuming a single eye-tracking to self-report mapping transfers uniformly across AI conditions and settings.


\begin{figure*}[!t]
    \centering
    \includegraphics[width=0.982\linewidth]{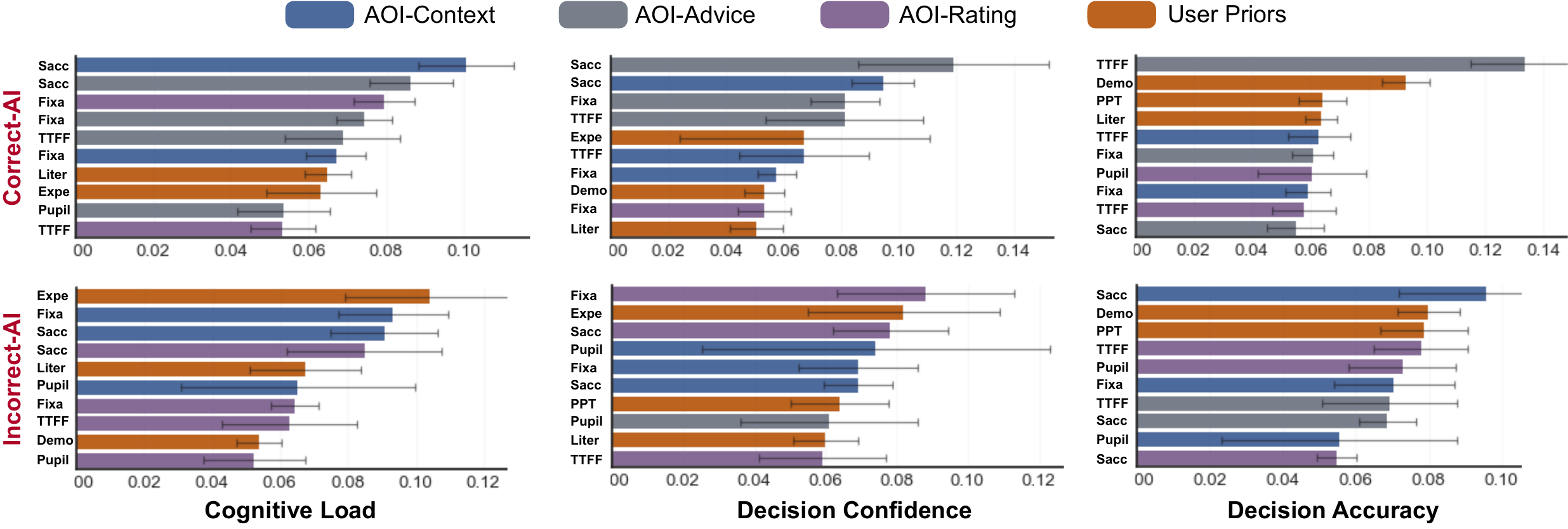}
    \vspace{-3.2mm}
    \caption{
    SHAP analysis: top 10 important features in user modeling on two AI conditions (Correct-AI vs. Incorrect-AI), to predict self-reported cognitive states (cognitive load and confidence) and decision accuracy by ExtraTree classifiers. 
    (``Sacc'' = saccades; ``Fixa'' = fixations; ``Pupil'' = Pupil Diameter; ``Liter'' = AI literacy; ``Expe'' = AI experience; ``Demo'' = Demographics;).
    }
    \label{fig:shap}
    \vspace{-1.8mm}
\end{figure*}


\vspace{-1.6mm}
\subsection{User modeling by multimodal fusion - \textit{RQ3}}

We next examine \textbf{RQ3} on whether fusing user priors improves predictive modeling beyond eye tracking alone. 
We compare three feature settings (Sec.~\ref{modeling}): 
\textit{eye-tracking only} ($g_i$), \textit{user priors only} ($p_{u_i}$), and \textit{multimodal fusion} (eye-tracking + user priors; $[g_i; p_{u_i}]$). 

\textit{Fusing priors with eye tracking yields the most consistent gains for \textit{cognitive load} and \textit{confidence} across conditions}, as shown in Table~\ref{tab:res_predictions} \& ~\ref{tab:res_predictions_best}.
Across predicted targets, multimodal fusion is competitive with or better than priors or gaze only, with the clearest gains for predicting \textbf{cognitive load} and \textbf{decision confidence}. 
For example, under pooled training (\textit{All}), fusion raises cognitive-load performance relative to eye-tracking only for several classifiers (e.g., MLP improves from \(.64/.74/.62\) to \(.73/.79/.76\)), indicating that user priors add useful information that is not captured by eye-tracking patterns alone. 
A similar pattern appears for decision confidence, where fusion reaches higher performance (e.g., XGBoost).
The gains of fusion are condition-dependent: improvements are typically larger when no advice is present than Correct-AI and Incorrect-AI, implying with the idea that user priors may complement how users attend to and utilize advice when misleading AI is introduced.

For decision accuracy, improvements from priors and fusion are smaller. This pattern suggests that while priors and eye-tracking signals are informative for modeling perceived cognitive states (\textit{load} and \textit{confidence}), predicting objective behaviors remains more challenging and may depend on factors beyond eye-tracking (e.g., AI experience, domain knowledge or item difficulty).

\textit{User priors are informative, but work best when combined with gaze.}
User priors only models already provide competitive predictive power, suggesting stable individual differences in self-report tendencies.
However, fusion is more consistently reliable across models and predicted targets than user priors alone, supporting a \emph{complementarity} view: priors capture between-participant tendencies, while gaze captures within-trial behavioral signals.

\textit{Where fusion helps most depends on the AI condition, revealing condition sensitivity.}
Best-performing settings concentrate on specific AI conditions rather than uniformly under pooled training. 
The summary of best settings (Table.~\ref{tab:res_predictions} \& ~\ref{tab:res_predictions_best}) shows a clear condition-dependent structure: (i) \textbf{decision confidence} often peaks under \textit{Correct-AI} with fusion (e.g., several models achieve \(\sim .74\) accuracy with strong F1/AUC), while (ii) \textbf{decision accuracy} consistently peaks under \textit{Incorrect-AI} with fusion (e.g., \(\sim .82{-}.85\) accuracy across multiple models). 
This indicates that priors are particularly helpful when the mapping from gaze to outcomes shifts with AI assistance, for instance, misleading AI can change strategies and uncertainty, making gaze alone less transferable across participants.


Overall, these findings address RQ3 that incorporating user priors improves user modeling when fused with eye-tracking, with gains that are \emph{condition-sensitive}.



\vspace{-1.2mm}
\subsection{Features importance across AI conditions}

Fig.~\ref{fig:shap} shows feature importance by SHAP~\citep{xai_shape} analysis for ExtraTrees models trained under \textit{Correct-AI} and \textit{Incorrect-AI} conditions. 

For \textbf{cognitive load}, \textit{Correct-AI} is dominated by eye-tracking-derived features (e.g., fixation/saccade-related metrics and pupil cues), suggesting that load is primarily reflected in visual processing dynamics when advice is reliable. However, under \textit{Incorrect-AI}, \textbf{user priors} (notably AI experience) rise sharply and can become top-ranked, implying that stable individual differences help explain effort variation when users face misleading assistance.

For \textbf{decision confidence}, features related to attention on \textbf{AI advice} and \textbf{rating} regions consistently appear among the top contributors, consistent with confidence being shaped by how users evaluate AI suggestions. 
Importantly, priors become more prominent under \textit{Incorrect-AI}, indicating that confidence under misleading advice are more strongly moderated by individual predispositions.

For \textbf{decision accuracy}, \textit{Correct-AI} shows a strong contribution of AOI-Advice feature, whereas \textit{Incorrect-AI} exhibits higher rankings on AOI-Context.
High reliance on \textbf{user priors} (demographics, literacy, trust propensity) supports the interpretation that priors act as an interpretable stabilizer for cross-participant generalization. 


\section{Discussion}

\subsection{Condition-sensitive user modeling}

A key takeaway from our study is that user state inference from eye-tracking is \emph{condition-sensitive}.
In our experiment, AI assistance systematically altered users' self-reported states (Table~\ref{tab:anova_pairwise}): relative to No-AI, both AI conditions increased decision confidence, and Correct-AI reduced perceived cognitive load. 
These shifts were accompanied by changes in eye-tracking patterns (RQ1) and translated into heterogeneous predictive performance across conditions (RQ2), indicating that the same sensing modality supports different levels of inference reliability depending on how AI is involved.

This condition sensitivity is also reflected in modeling outcomes. 
Training on pooled trials from all conditions did not consistently outperform condition-specific modeling, and the strongest results for confidence and cognitive load often emerged from specific contexts rather than the pooled set (Table~\ref{tab:res_predictions}). 
A plausible mechanism is that AI involvement changes users' decision strategy and attention allocation~\cite{sun_trust,7451741}: when AI suggestions are present, users may skim evidence differently, rely more on the advice panel, or experience lower effort during rating; when suggestions are misleading, users may experience conflict and adopt different verification patterns (Fig.~\ref{fig:lab_gaze_analysis}).
Such strategy shifts align with prior human-AI interaction findings that perceived reliability, trust, and transparency affect how users engage with AI outputs and when they rely on suggestions~\cite{trust_ca,appropriate_reliance_AI_advic,explanation_ai_overreliance,Wang2023-nm,Vereschak2024,incorrect_xai_1}. 
Our findings show that these condition-induced shifts are not only reflected in cognitive outcomes (e.g., confidence/load) but are also observable in implicit gaze behaviors and affect the robustness of eye-tracking-based user models.

Importantly, condition sensitivity does not mean modeling by eye-tracking is ineffective; rather, it implies that \emph{AI conditions should be treated as a feature in user modeling}.
Our condition-stratified testing demonstrates that even with the same eye-tracking features, model performance varies across AI conditions and differs by predicted targets (Table~\ref{tab:res_predictions}). 
This finding echoes prior work using eye tracking to infer cognitive effort in stable tasks~\cite{eyetracking_survey,physio_search,eyetracking_cognitive_2,Parikh2018EyeGF,SUMER2021106909} by highlighting a critical boundary condition for deployable user models: when the interaction context itself shifts due to AI assistance, signal attribution and learned mappings can shift as well.

\vspace{-1.6mm}
\paragraph{\textbf{Practical implication.}} 
For deployable user models, evaluation and deployment should be \emph{condition-aware}. 
At minimum, models should report condition-stratified performance (rather than only overall accuracy) and incorporate observable AI-condition features (e.g., assistance mode, presence/format of suggestions or explanations, correctness) to avoid averaging over heterogeneous behaviors. 
Where condition heterogeneity is strong, simple condition-aware architectures (e.g., condition-specific heads or expert routing) can provide a robust alternative to a single pooled estimator. 

Beyond improving predictive user modeling, our results motivate a design for \emph{cognitively-aligned} AI assistance: systems should jointly model (i) moment-to-moment attention and cognitive states from behavioral signals, (ii) AI conditions (e.g., assistance mode and reliability cues), and (iii) stable individual differences (e.g., AI literacy) to decide \emph{when}, \emph{how}, and \emph{for whom} to adapt assistance. 
Concretely, this suggests a closed-loop pipeline in which eye-tracking provides continuous state estimates, condition-aware modeling handles condition shifts (e.g., via condition features or expert routing), and user priors provide personalized capability to the systems. 

These design recommendations are consistent with prior HCI practices that reliability and trust to AI influence overreliance, engagement and user behaviors~\cite{explanation_ai_overreliance,Wang2023-nm,Vereschak2024} and with sensing-based HCI work cautioning that behavioral and physiological signals require careful attribution under changing task demands and real-time inference for adaptive interaction~\cite{sun_trust,Chiossi2024,context_recsys_1,context_recsys_2}.


\vspace{-1.0mm}
\subsection{User priors for cold-start generalization in user modeling from eye-tracking}

Our results further show that incorporating stable user priors, demographics, AI literacy, and propensity trust in technology, improves the \emph{cross-participant} generalization of eye-tracking-based user modeling (RQ3). 
Under participant-level evaluation, multimodal fusion (eye-tracking + priors) consistently outperforms eye-tracking-only models (Table~\ref{tab:res_predictions}), with the clearest gains in accuracy/F1 and threshold-independent discrimination (AUC), even when accuracy changes are smaller. 
This pattern suggests that user priors help \emph{calibrate and rank} users' subjective states rather than merely shifting decisions across a fixed classification threshold.

These gains are particularly meaningful given the gap between trial-level and participant-level evaluation. 
Trial-level $k$-fold cross-validation is often optimistic because models can leverage stable, person-specific baselines in eye-tracking (e.g., reading speed, fixation style, pupil dynamics) when the same individual appears in both train and test splits. 
Our findings indicate that user priors provide an interpretable anchor for the \emph{cold-start} setting~\cite{cold_start} by reducing cross-user ambiguity, i.e., they help explain why the \emph{same} eye-tracking pattern may correspond to different reported confidence or cognitive load across individuals. 
This is consistent with prior human-AI interaction research showing that individual differences such as AI literacy and trust level influence how people interpret and rely on AI systems or advice~\cite{Vereschak2024,explanation_ai_overreliance}, and extends that line of work by demonstrating a concrete modeling benefit when these traits are used as priors for user cognitive state estimation.

Importantly, the improvements are not uniform across conditions. 
Condition-stratified tests show that fusion benefits concentrate in AI conditions with stronger heterogeneity (Table~\ref{tab:res_predictions}), where eyetracking-states mappings are less stable and pooled modeling is more likely to average over divergent strategies. 
Thus, user priors function as a stabilizer that compensates for condition-induced variability by encoding stable tendencies (e.g., baseline experience and literacy level) that modulate how eye-tracking relates to subjective effort and certainty. 
This finding also helps interpret the earlier observation that ``pooled'' (all) training does not always help (RQ2).

\vspace{-2.2mm}
\paragraph{\textbf{Practical implication.}} 
Our results suggest treating user priors as a principled component of cold-start state inference rather than an optional add-on. 
In practice, systems and user models can obtain lightweight priors via short onboarding questionnaires or inferred proxies (e.g., interaction history) and fuse them with behavioral sensing to improve generalization for new users. 
This recommendation is grounded in prior HAI evidence that trust and literacy affect reliance and overreliance behaviors and aligns with user modeling principles that stable user traits can complement short-term behavioral signals to support personalized user modeling.

\section{Limitations and Future Work}
We acknowledge several limitations of this work as follows.

\textbf{First, self-reports as labels.} 
We model self-reported cognitive load and confidence, which are subjective and may be noisy due to individual differences. We used a brief load scale instead of NASA-TLX~\cite{NASA_cognitive_load} to reduce fatigue. 
\textit{Future work:} We should triangulate self-reports with complementary indicators (e.g., response time) and improve reliability via repeated ratings. 
We should caution against treating eye-tracking as a causal proxy for cognitive states~\citep{Cacioppo}.

\textbf{Second, task specificity.} 
We study StrategyQA-style factual verification, which offers control but may not represent broader AI-assisted decision-making. 
Our main conclusions about context shifts and the value of user priors require replication in other tasks and stakes. 
\textit{Future work:} We will replicate the study across decision types and test whether the same condition shifts and personalization gains hold under varied cognitive demands and stakes.

\textbf{Third, fixed AI advice format.} 
AI advice is a brief suggestion and incorrect advice is created by flipping the ground-truth. 
This does not cover richer AI assistance formats (e.g., explanations). 
\textit{Future work:} We will vary assistance style and introduce realistic errors (partial correctness, misleading evidence, hallucinations~\cite{Hallucin28:online}).


\textbf{Lastly, sample size and participant diversity.} 
Our participant pool may limit the power of generalized user modeling. 
Nonetheless, our primary findings rely on within-subject comparisons, which support the robustness of the observed condition sensitivity and fusion gains. 
\textit{Future work:} We will scale to larger and more diverse samples to test stability across demographic groups. 


\section{Conclusion}


We investigated user modeling on cognitive states from eye-tracking signals and user priors in AI-assisted decision-making across three AI conditions (No-AI, Correct-AI, Incorrect-AI). 
The findings show that AI advice affects both self-reported cognitive states, decision-making and eye-tracking patterns. 
We further reveal that user modeling from eye-tracking alone is condition-sensitive, and fusing eye-tracking with user priors (demographics, AI literacy/experience, and propensity trust) can improve cross-participant generalization for user modeling. 
Overall, our work suggests that cognitive-state user models should incorporate both AI conditions and individual differences to enable cognitively-aligned adaptive AI systems.

\section*{GenAI Declaration}

We used AI tools (i.e., GPT-5.2) only for language editing to improve clarity and conciseness.
All study design, analysis, literature review, and writing were conducted and verified by the authors.



\begin{acks}
This work was supported by the JST CREST Grant (JPMJCR2562), JST K Program Grant (JPMJKP24C2), and JST FOREST Grant (JPMJFR232R) in Japan.
\end{acks}






\twocolumn

\bibliographystyle{ACM-Reference-Format}
\bibliography{references}


\begin{thebibliography}{61}


\ifx \showCODEN    \undefined \def \showCODEN     #1{\unskip}     \fi
\ifx \showDOI      \undefined \def \showDOI       #1{#1}\fi
\ifx \showISBNx    \undefined \def \showISBNx     #1{\unskip}     \fi
\ifx \showISBNxiii \undefined \def \showISBNxiii  #1{\unskip}     \fi
\ifx \showISSN     \undefined \def \showISSN      #1{\unskip}     \fi
\ifx \showLCCN     \undefined \def \showLCCN      #1{\unskip}     \fi
\ifx \shownote     \undefined \def \shownote      #1{#1}          \fi
\ifx \showarticletitle \undefined \def \showarticletitle #1{#1}   \fi
\ifx \showURL      \undefined \def \showURL       {\relax}        \fi
\providecommand\bibfield[2]{#2}
\providecommand\bibinfo[2]{#2}
\providecommand\natexlab[1]{#1}
\providecommand\showeprint[2][]{arXiv:#2}

\bibitem[mix(1988)]%
        {mixedLM}
 \bibinfo{year}{1988}\natexlab{}.
\newblock \showarticletitle{Newton Raphson and {EM} algorithms for linear mixed
  effects models for repeated measures data}.
\newblock \bibinfo{journal}{\emph{J. Amer. Statist. Assoc.}}
  \bibinfo{volume}{83}, \bibinfo{number}{404} (\bibinfo{year}{1988}),
  \bibinfo{pages}{1014--1022}.
\newblock


\bibitem[Abdrabou et~al\mbox{.}(2023)]%
        {Abdrabou_Yasmeen}
\bibfield{author}{\bibinfo{person}{Yasmeen Abdrabou},
  \bibinfo{person}{Elisaveta Karypidou}, \bibinfo{person}{Florian Alt}, {and}
  \bibinfo{person}{Mariam Hassib}.} \bibinfo{year}{2023}\natexlab{}.
\newblock \showarticletitle{Investigating User Behavior Towards Fake News on
  Social Media Using Gaze and Mouse Movements}.
\newblock
\urldef\tempurl%
\url{https://doi.org/10.14722/usec.2023.232041}
\showDOI{\tempurl}


\bibitem[Ahmad and Alzahrani(2023)]%
        {Ahmad_Muneeb}
\bibfield{author}{\bibinfo{person}{Muneeb Ahmad} {and}
  \bibinfo{person}{Abdullah Alzahrani}.} \bibinfo{year}{2023}\natexlab{}.
\newblock \showarticletitle{Crucial Clues: Investigating Psychophysiological
  Behaviors for Measuring Trust in Human-Robot Interaction}. In
  \bibinfo{booktitle}{\emph{Proceedings of the 25th International Conference on
  Multimodal Interaction}} (Paris, France) \emph{(\bibinfo{series}{ICMI '23})}.
  \bibinfo{publisher}{Association for Computing Machinery},
  \bibinfo{address}{New York, NY, USA}, \bibinfo{pages}{135–143}.
\newblock
\showISBNx{9798400700552}
\urldef\tempurl%
\url{https://doi.org/10.1145/3577190.3614148}
\showDOI{\tempurl}


\bibitem[Ajenaghughrure et~al\mbox{.}(2021)]%
        {Ajenaghughrure_modeling}
\bibfield{author}{\bibinfo{person}{Ighoyota~Ben. Ajenaghughrure},
  \bibinfo{person}{Sònia~Cláudia Da~Costa~Sousa}, {and}
  \bibinfo{person}{David Lamas}.} \bibinfo{year}{2021}\natexlab{}.
\newblock \showarticletitle{Psychophysiological modelling of trust in
  technology: Comparative analysis of algorithm ensemble methods}. In
  \bibinfo{booktitle}{\emph{2021 IEEE 19th World Symposium on Applied Machine
  Intelligence and Informatics (SAMI)}}. \bibinfo{pages}{000161--000168}.
\newblock
\urldef\tempurl%
\url{https://doi.org/10.1109/SAMI50585.2021.9378655}
\showDOI{\tempurl}


\bibitem[Akash et~al\mbox{.}(2018)]%
        {Akash_Kumar}
\bibfield{author}{\bibinfo{person}{Kumar Akash}, \bibinfo{person}{Wan-Lin Hu},
  \bibinfo{person}{Neera Jain}, {and} \bibinfo{person}{Tahira Reid}.}
  \bibinfo{year}{2018}\natexlab{}.
\newblock \showarticletitle{A Classification Model for Sensing Human Trust in
  Machines Using EEG and GSR}.
\newblock \bibinfo{journal}{\emph{ACM Trans. Interact. Intell. Syst.}}
  \bibinfo{volume}{8}, \bibinfo{number}{4}, Article \bibinfo{articleno}{27}
  (\bibinfo{date}{nov} \bibinfo{year}{2018}), \bibinfo{numpages}{20}~pages.
\newblock
\showISSN{2160-6455}
\urldef\tempurl%
\url{https://doi.org/10.1145/3132743}
\showDOI{\tempurl}


\bibitem[Bachmann(2025)]%
        {xai_shape}
\bibfield{author}{\bibinfo{person}{Severin Bachmann}.}
  \bibinfo{year}{2025}\natexlab{}.
\newblock \showarticletitle{Efficient XAI: A Low-Cost Data Reduction Approach
  to SHAP Interpretability}.
\newblock \bibinfo{journal}{\emph{J. Artif. Int. Res.}} (\bibinfo{year}{2025}),
  \bibinfo{numpages}{21}~pages.
\newblock
\showISSN{1076-9757}
\urldef\tempurl%
\url{https://doi.org/10.1613/jair.1.18325}
\showDOI{\tempurl}


\bibitem[Boonprakong et~al\mbox{.}(2023)]%
        {eyetracking_cognitive_1}
\bibfield{author}{\bibinfo{person}{Nattapat Boonprakong},
  \bibinfo{person}{Xiuge Chen}, \bibinfo{person}{Catherine Davey},
  \bibinfo{person}{Benjamin Tag}, {and} \bibinfo{person}{Tilman Dingler}.}
  \bibinfo{year}{2023}\natexlab{}.
\newblock \showarticletitle{Bias-Aware Systems: Exploring Indicators for the
  Occurrences of Cognitive Biases when Facing Different Opinions}. In
  \bibinfo{booktitle}{\emph{Proceedings of the 2023 CHI Conference on Human
  Factors in Computing Systems}} (Hamburg, Germany) \emph{(\bibinfo{series}{CHI
  '23})}. \bibinfo{publisher}{Association for Computing Machinery},
  \bibinfo{address}{New York, NY, USA}, Article \bibinfo{articleno}{27},
  \bibinfo{numpages}{19}~pages.
\newblock
\showISBNx{9781450394215}
\urldef\tempurl%
\url{https://doi.org/10.1145/3544548.3580917}
\showDOI{\tempurl}


\bibitem[Cacioppo et~al\mbox{.}(2016)]%
        {Cacioppo}
\bibfield{author}{\bibinfo{person}{John~T. Cacioppo}, \bibinfo{person}{Louis~G.
  Tassinary}, {and} \bibinfo{person}{Gary~G. Berntson}.}
  \bibinfo{year}{2016}\natexlab{}.
\newblock \bibinfo{booktitle}{\emph{Strong Inference in Psychophysiological
  Science}}.
\newblock \bibinfo{publisher}{Cambridge University Press},
  \bibinfo{pages}{3–15}.
\newblock


\bibitem[Carolus et~al\mbox{.}(2023)]%
        {ai_literacy}
\bibfield{author}{\bibinfo{person}{Astrid Carolus}, \bibinfo{person}{Martin~J.
  Koch}, \bibinfo{person}{Samantha Straka}, \bibinfo{person}{Marc~Erich
  Latoschik}, {and} \bibinfo{person}{Carolin Wienrich}.}
  \bibinfo{year}{2023}\natexlab{}.
\newblock \showarticletitle{MAILS - Meta AI literacy scale: Development and
  testing of an AI literacy questionnaire based on well-founded competency
  models and psychological change- and meta-competencies}.
\newblock \bibinfo{journal}{\emph{Computers in Human Behavior: Artificial
  Humans}} \bibinfo{volume}{1}, \bibinfo{number}{2} (\bibinfo{year}{2023}),
  \bibinfo{pages}{100014}.
\newblock
\showISSN{2949-8821}
\urldef\tempurl%
\url{https://doi.org/10.1016/j.chbah.2023.100014}
\showDOI{\tempurl}


\bibitem[Chiossi et~al\mbox{.}(2024)]%
        {Chiossi2024}
\bibfield{author}{\bibinfo{person}{Francesco Chiossi},
  \bibinfo{person}{Ekaterina~R. Stepanova}, \bibinfo{person}{Benjamin Tag},
  \bibinfo{person}{Monica Perusquia-Hernandez}, \bibinfo{person}{Alexandra
  Kitson}, \bibinfo{person}{Arindam Dey}, \bibinfo{person}{Sven Mayer}, {and}
  \bibinfo{person}{Abdallah El~Ali}.} \bibinfo{year}{2024}\natexlab{}.
\newblock \showarticletitle{PhysioCHI: Towards Best Practices for Integrating
  Physiological Signals in HCI}. In \bibinfo{booktitle}{\emph{Extended
  Abstracts of the CHI Conference on Human Factors in Computing Systems}}
  (Honolulu, HI, USA) \emph{(\bibinfo{series}{CHI EA '24})}.
  \bibinfo{publisher}{Association for Computing Machinery},
  \bibinfo{address}{New York, NY, USA}, Article \bibinfo{articleno}{485},
  \bibinfo{numpages}{7}~pages.
\newblock
\showISBNx{9798400703317}
\urldef\tempurl%
\url{https://doi.org/10.1145/3613905.3636286}
\showDOI{\tempurl}


\bibitem[Christen et~al\mbox{.}(2023)]%
        {f1}
\bibfield{author}{\bibinfo{person}{Peter Christen}, \bibinfo{person}{David~J.
  Hand}, {and} \bibinfo{person}{Nishadi Kirielle}.}
  \bibinfo{year}{2023}\natexlab{}.
\newblock \showarticletitle{A Review of the F-Measure: Its History, Properties,
  Criticism, and Alternatives}.
\newblock \bibinfo{journal}{\emph{ACM Comput. Surv.}} \bibinfo{volume}{56},
  \bibinfo{number}{3}, Article \bibinfo{articleno}{73} (\bibinfo{date}{Oct.}
  \bibinfo{year}{2023}), \bibinfo{numpages}{24}~pages.
\newblock
\showISSN{0360-0300}
\urldef\tempurl%
\url{https://doi.org/10.1145/3606367}
\showDOI{\tempurl}


\bibitem[Faul et~al\mbox{.}(2007)]%
        {gpower}
\bibfield{author}{\bibinfo{person}{Franz Faul}, \bibinfo{person}{Edgar
  Erdfelder}, \bibinfo{person}{Albert-Georg Lang}, {and} \bibinfo{person}{Axel
  Buchner}.} \bibinfo{year}{2007}\natexlab{}.
\newblock \showarticletitle{{G*Power} 3: a flexible statistical power analysis
  program for the social, behavioral, and biomedical sciences}.
\newblock \bibinfo{journal}{\emph{Behav. Res. Methods}} \bibinfo{volume}{39},
  \bibinfo{number}{2} (\bibinfo{date}{May} \bibinfo{year}{2007}),
  \bibinfo{pages}{175--191}.
\newblock


\bibitem[Geva et~al\mbox{.}(2021)]%
        {strategyqa}
\bibfield{author}{\bibinfo{person}{Mor Geva}, \bibinfo{person}{Daniel
  Khashabi}, \bibinfo{person}{Elad Segal}, \bibinfo{person}{Tushar Khot},
  \bibinfo{person}{Dan Roth}, {and} \bibinfo{person}{Jonathan Berant}.}
  \bibinfo{year}{2021}\natexlab{}.
\newblock \bibinfo{title}{Did Aristotle Use a Laptop? A Question Answering
  Benchmark with Implicit Reasoning Strategies}.
\newblock
\newblock
\showeprint[arxiv]{2101.02235}~[cs.CL]
\urldef\tempurl%
\url{https://arxiv.org/abs/2101.02235}
\showURL{%
\tempurl}


\bibitem[Ghosh et~al\mbox{.}(2025)]%
        {context_recsys_2}
\bibfield{author}{\bibinfo{person}{Indrajeet Ghosh}, \bibinfo{person}{Kasthuri
  Jayarajah}, \bibinfo{person}{Nicholas Waytowich}, {and}
  \bibinfo{person}{Nirmalya Roy}.} \bibinfo{year}{2025}\natexlab{}.
\newblock \showarticletitle{Augmenting Personalized Memory via Practical
  Multimodal Wearable Sensing in Visual Search and Wayfinding Navigation}. In
  \bibinfo{booktitle}{\emph{Proceedings of the 33rd ACM Conference on User
  Modeling, Adaptation and Personalization}} \emph{(\bibinfo{series}{UMAP
  '25})}. \bibinfo{publisher}{Association for Computing Machinery},
  \bibinfo{address}{New York, NY, USA}, \bibinfo{pages}{11–21}.
\newblock
\showISBNx{9798400713132}
\urldef\tempurl%
\url{https://doi.org/10.1145/3699682.3728340}
\showDOI{\tempurl}


\bibitem[Girden(1992)]%
        {anova}
\bibfield{author}{\bibinfo{person}{Ellen~R Girden}.}
  \bibinfo{year}{1992}\natexlab{}.
\newblock \bibinfo{booktitle}{\emph{ANOVA: Repeated measures}}.
\newblock Number~84. \bibinfo{publisher}{Sage}.
\newblock


\bibitem[Gomez et~al\mbox{.}(2025)]%
        {AI_assisted_decision_making}
\bibfield{author}{\bibinfo{person}{Catalina Gomez}, \bibinfo{person}{Sue~Min
  Cho}, \bibinfo{person}{Shichang Ke}, \bibinfo{person}{Chien-Ming Huang},
  {and} \bibinfo{person}{Mathias Unberath}.} \bibinfo{year}{2025}\natexlab{}.
\newblock \showarticletitle{Human-AI collaboration is not very collaborative
  yet: a taxonomy of interaction patterns in AI-assisted decision making from a
  systematic review}.
\newblock \bibinfo{journal}{\emph{Frontiers in Computer Science}}
  \bibinfo{volume}{Volume 6 - 2024} (\bibinfo{year}{2025}).
\newblock
\showISSN{2624-9898}
\urldef\tempurl%
\url{https://doi.org/10.3389/fcomp.2024.1521066}
\showDOI{\tempurl}


\bibitem[Gupta et~al\mbox{.}(2022)]%
        {trust_ca}
\bibfield{author}{\bibinfo{person}{Akshit Gupta}, \bibinfo{person}{Debadeep
  Basu}, \bibinfo{person}{Ramya Ghantasala}, \bibinfo{person}{Sihang Qiu},
  {and} \bibinfo{person}{Ujwal Gadiraju}.} \bibinfo{year}{2022}\natexlab{}.
\newblock \showarticletitle{To Trust or Not To Trust: How a Conversational
  Interface Affects Trust in a Decision Support System}. In
  \bibinfo{booktitle}{\emph{Proceedings of the ACM Web Conference 2022}}
  (Virtual Event, Lyon, France) \emph{(\bibinfo{series}{WWW '22})}.
  \bibinfo{publisher}{Association for Computing Machinery},
  \bibinfo{address}{New York, NY, USA}, \bibinfo{pages}{3531–3540}.
\newblock
\showISBNx{9781450390965}
\urldef\tempurl%
\url{https://doi.org/10.1145/3485447.3512248}
\showDOI{\tempurl}


\bibitem[Hart and Staveland(1988)]%
        {NASA_cognitive_load}
\bibfield{author}{\bibinfo{person}{Sandra~G Hart} {and}
  \bibinfo{person}{Lowell~E Staveland}.} \bibinfo{year}{1988}\natexlab{}.
\newblock \showarticletitle{Development of {NASA-TLX} (task load index):
  Results of empirical and theoretical research}.
\newblock In \bibinfo{booktitle}{\emph{Advances in Psychology}}.
  \bibinfo{publisher}{Elsevier}, \bibinfo{pages}{139--183}.
\newblock


\bibitem[Haynes(2013)]%
        {fdr_bh}
\bibfield{author}{\bibinfo{person}{Winston Haynes}.}
  \bibinfo{year}{2013}\natexlab{}.
\newblock \bibinfo{booktitle}{\emph{Benjamini--Hochberg Method}}.
\newblock \bibinfo{publisher}{Springer New York}, \bibinfo{address}{New York,
  NY}, \bibinfo{pages}{78--78}.
\newblock
\showISBNx{978-1-4419-9863-7}
\urldef\tempurl%
\url{https://doi.org/10.1007/978-1-4419-9863-7\_1215}
\showDOI{\tempurl}


\bibitem[Hoffman et~al\mbox{.}(2013)]%
        {measure_trust_automation}
\bibfield{author}{\bibinfo{person}{Robert~R. Hoffman}, \bibinfo{person}{Matthew
  Johnson}, \bibinfo{person}{Jeffrey~M. Bradshaw}, {and} \bibinfo{person}{Al
  Underbrink}.} \bibinfo{year}{2013}\natexlab{}.
\newblock \showarticletitle{Trust in Automation}.
\newblock \bibinfo{journal}{\emph{IEEE Intelligent Systems}}
  \bibinfo{volume}{28}, \bibinfo{number}{1} (\bibinfo{date}{Jan.}
  \bibinfo{year}{2013}), \bibinfo{pages}{84–88}.
\newblock
\showISSN{1541-1672}
\urldef\tempurl%
\url{https://doi.org/10.1109/MIS.2013.24}
\showDOI{\tempurl}


\bibitem[Holmqvist et~al\mbox{.}(2011)]%
        {eyetracking_methods}
\bibfield{author}{\bibinfo{person}{Kenneth Holmqvist}, \bibinfo{person}{Marcus
  Nystrom}, \bibinfo{person}{Richard Andersson}, \bibinfo{person}{Richard
  Dewhurst}, \bibinfo{person}{Halszka Jarodzka}, {and} \bibinfo{person}{Joost
  Van~de Weijer}.} \bibinfo{year}{2011}\natexlab{}.
\newblock \bibinfo{booktitle}{\emph{Eye tracking: A comprehensive guide to
  methods and measures}}.
\newblock \bibinfo{publisher}{Oxford University Press},
  \bibinfo{address}{United States}.
\newblock
\showISBNx{978-0-19969708-3}


\bibitem[Jessup et~al\mbox{.}(2019)]%
        {ppt}
\bibfield{author}{\bibinfo{person}{Sarah Jessup}, \bibinfo{person}{Tamera
  Schneider}, \bibinfo{person}{Gene Alarcon}, \bibinfo{person}{Tyler Ryan},
  {and} \bibinfo{person}{August Capiola}.} \bibinfo{year}{2019}\natexlab{}.
\newblock \bibinfo{booktitle}{\emph{The Measurement of the Propensity to Trust
  Automation}}.
\newblock \bibinfo{pages}{476--489}.
\newblock
\showISBNx{978-3-030-21564-4}
\urldef\tempurl%
\url{https://doi.org/10.1007/978-3-030-21565-1_32}
\showDOI{\tempurl}


\bibitem[Ji et~al\mbox{.}(2024)]%
        {physio_search}
\bibfield{author}{\bibinfo{person}{Kaixin Ji}, \bibinfo{person}{Danula
  Hettiachchi}, \bibinfo{person}{Flora~D. Salim}, \bibinfo{person}{Falk
  Scholer}, {and} \bibinfo{person}{Damiano Spina}.}
  \bibinfo{year}{2024}\natexlab{}.
\newblock \showarticletitle{Characterizing Information Seeking Processes with
  Multiple Physiological Signals}. In \bibinfo{booktitle}{\emph{Proceedings of
  the 47th International ACM SIGIR Conference on Research and Development in
  Information Retrieval}} \emph{(\bibinfo{series}{SIGIR 2024},
  Vol.~\bibinfo{volume}{5})}. \bibinfo{publisher}{ACM},
  \bibinfo{pages}{1006–1017}.
\newblock
\urldef\tempurl%
\url{https://doi.org/10.1145/3626772.3657793}
\showDOI{\tempurl}


\bibitem[Jin et~al\mbox{.}(2019)]%
        {context_recsys_1}
\bibfield{author}{\bibinfo{person}{Yucheng Jin}, \bibinfo{person}{Nyi~Nyi
  Htun}, \bibinfo{person}{Nava Tintarev}, {and} \bibinfo{person}{Katrien
  Verbert}.} \bibinfo{year}{2019}\natexlab{}.
\newblock \showarticletitle{ContextPlay: Evaluating User Control for
  Context-Aware Music Recommendation}. In \bibinfo{booktitle}{\emph{Proceedings
  of the 27th ACM Conference on User Modeling, Adaptation and Personalization}}
  (Larnaca, Cyprus) \emph{(\bibinfo{series}{UMAP '19})}.
  \bibinfo{publisher}{Association for Computing Machinery},
  \bibinfo{address}{New York, NY, USA}, \bibinfo{pages}{294–302}.
\newblock
\showISBNx{9781450360210}
\urldef\tempurl%
\url{https://doi.org/10.1145/3320435.3320445}
\showDOI{\tempurl}


\bibitem[Kunjan et~al\mbox{.}(2021)]%
        {loso}
\bibfield{author}{\bibinfo{person}{Sajeev Kunjan}, \bibinfo{person}{T.~S.
  Grummett}, \bibinfo{person}{K.~J. Pope}, \bibinfo{person}{D.~M.~W. Powers},
  \bibinfo{person}{S.~P. Fitzgibbon}, \bibinfo{person}{T. Bastiampillai},
  \bibinfo{person}{M. Battersby}, {and} \bibinfo{person}{T.~W. Lewis}.}
  \bibinfo{year}{2021}\natexlab{}.
\newblock \showarticletitle{The Necessity of Leave One Subject Out (LOSO) Cross
  Validation for EEG Disease Diagnosis}. In \bibinfo{booktitle}{\emph{Brain
  Informatics: 14th International Conference, BI 2021, Virtual Event, September
  17–19, 2021, Proceedings}}. \bibinfo{publisher}{Springer-Verlag},
  \bibinfo{address}{Berlin, Heidelberg}, \bibinfo{pages}{558–567}.
\newblock
\showISBNx{978-3-030-86992-2}
\urldef\tempurl%
\url{https://doi.org/10.1007/978-3-030-86993-9_50}
\showDOI{\tempurl}


\bibitem[Kunkel et~al\mbox{.}(2019)]%
        {10.1145/3290605.3300717}
\bibfield{author}{\bibinfo{person}{Johannes Kunkel}, \bibinfo{person}{Tim
  Donkers}, \bibinfo{person}{Lisa Michael}, \bibinfo{person}{Catalin-Mihai
  Barbu}, {and} \bibinfo{person}{J\"{u}rgen Ziegler}.}
  \bibinfo{year}{2019}\natexlab{}.
\newblock \showarticletitle{Let Me Explain: Impact of Personal and Impersonal
  Explanations on Trust in Recommender Systems}. In
  \bibinfo{booktitle}{\emph{Proceedings of the 2019 CHI Conference on Human
  Factors in Computing Systems}} (Glasgow, Scotland Uk)
  \emph{(\bibinfo{series}{CHI '19})}. \bibinfo{publisher}{Association for
  Computing Machinery}, \bibinfo{address}{New York, NY, USA},
  \bibinfo{pages}{1–12}.
\newblock
\showISBNx{9781450359702}
\urldef\tempurl%
\url{https://doi.org/10.1145/3290605.3300717}
\showDOI{\tempurl}


\bibitem[Lakkaraju and Bastani(2020)]%
        {10.1145/3375627.3375833}
\bibfield{author}{\bibinfo{person}{Himabindu Lakkaraju} {and}
  \bibinfo{person}{Osbert Bastani}.} \bibinfo{year}{2020}\natexlab{}.
\newblock \showarticletitle{"How do I fool you?": Manipulating User Trust via
  Misleading Black Box Explanations}. In \bibinfo{booktitle}{\emph{Proceedings
  of the AAAI/ACM Conference on AI, Ethics, and Society}} (New York, NY, USA)
  \emph{(\bibinfo{series}{AIES '20})}. \bibinfo{publisher}{Association for
  Computing Machinery}, \bibinfo{address}{New York, NY, USA},
  \bibinfo{pages}{79–85}.
\newblock
\showISBNx{9781450371100}
\urldef\tempurl%
\url{https://doi.org/10.1145/3375627.3375833}
\showDOI{\tempurl}


\bibitem[Leiser et~al\mbox{.}(2023)]%
        {Hallucinations}
\bibfield{author}{\bibinfo{person}{Florian Leiser}, \bibinfo{person}{Sven
  Eckhardt}, \bibinfo{person}{Merlin Knaeble}, \bibinfo{person}{Alexander
  Maedche}, \bibinfo{person}{Gerhard Schwabe}, {and} \bibinfo{person}{Ali
  Sunyaev}.} \bibinfo{year}{2023}\natexlab{}.
\newblock \showarticletitle{From ChatGPT to FactGPT: A Participatory Design
  Study to Mitigate the Effects of Large Language Model Hallucinations on
  Users}. In \bibinfo{booktitle}{\emph{Proceedings of Mensch Und Computer
  2023}} (Rapperswil, Switzerland) \emph{(\bibinfo{series}{MuC '23})}.
  \bibinfo{publisher}{Association for Computing Machinery},
  \bibinfo{address}{New York, NY, USA}, \bibinfo{pages}{81–90}.
\newblock
\showISBNx{9798400707711}
\urldef\tempurl%
\url{https://doi.org/10.1145/3603555.3603565}
\showDOI{\tempurl}


\bibitem[Li et~al\mbox{.}(2025)]%
        {measure_confidence}
\bibfield{author}{\bibinfo{person}{Jingshu Li}, \bibinfo{person}{Yitian Yang},
  \bibinfo{person}{Q.~Vera Liao}, \bibinfo{person}{Junti Zhang}, {and}
  \bibinfo{person}{Yi-Chieh Lee}.} \bibinfo{year}{2025}\natexlab{}.
\newblock \showarticletitle{As Confidence Aligns: Understanding the Effect of
  AI Confidence on Human Self-confidence in Human-AI Decision Making}. In
  \bibinfo{booktitle}{\emph{Proceedings of the 2025 CHI Conference on Human
  Factors in Computing Systems}} \emph{(\bibinfo{series}{CHI '25})}.
  \bibinfo{publisher}{ACM}, \bibinfo{address}{New York, NY, USA}, Article
  \bibinfo{articleno}{1111}, \bibinfo{numpages}{16}~pages.
\newblock
\showISBNx{9798400713941}
\urldef\tempurl%
\url{https://doi.org/10.1145/3706598.3713336}
\showDOI{\tempurl}


\bibitem[Liao and Sundar(2022)]%
        {responsible_ai}
\bibfield{author}{\bibinfo{person}{Q.Vera Liao} {and} \bibinfo{person}{S.~Shyam
  Sundar}.} \bibinfo{year}{2022}\natexlab{}.
\newblock \showarticletitle{Designing for Responsible Trust in AI Systems: A
  Communication Perspective}. In \bibinfo{booktitle}{\emph{Proceedings of the
  2022 ACM Conference on Fairness, Accountability, and Transparency}} (Seoul,
  Republic of Korea) \emph{(\bibinfo{series}{FAccT '22})}.
  \bibinfo{publisher}{ACM}, \bibinfo{address}{New York, NY, USA},
  \bibinfo{pages}{1257–1268}.
\newblock
\showISBNx{9781450393522}
\urldef\tempurl%
\url{https://doi.org/10.1145/3531146.3533182}
\showDOI{\tempurl}


\bibitem[Liao et~al\mbox{.}(2024)]%
        {ux_responsible}
\bibfield{author}{\bibinfo{person}{Q.~Vera Liao}, \bibinfo{person}{Mihaela
  Vorvoreanu}, \bibinfo{person}{Hari Subramonyam}, {and}
  \bibinfo{person}{Lauren Wilcox}.} \bibinfo{year}{2024}\natexlab{}.
\newblock \showarticletitle{UX Matters: The Critical Role of UX in Responsible
  AI}.
\newblock \bibinfo{journal}{\emph{Interactions}} \bibinfo{volume}{31},
  \bibinfo{number}{4} (\bibinfo{date}{jun} \bibinfo{year}{2024}),
  \bibinfo{pages}{22–27}.
\newblock
\showISSN{1072-5520}
\urldef\tempurl%
\url{https://doi.org/10.1145/3665504}
\showDOI{\tempurl}


\bibitem[Lim et~al\mbox{.}(2022)]%
        {eyetracking_cognitive_2}
\bibfield{author}{\bibinfo{person}{Jia~Zheng Lim}, \bibinfo{person}{James
  Mountstephens}, {and} \bibinfo{person}{Jason Teo}.}
  \bibinfo{year}{2022}\natexlab{}.
\newblock \showarticletitle{Eye-Tracking Feature Extraction for Biometric
  Machine Learning}.
\newblock \bibinfo{journal}{\emph{Front Neurorobot}}  \bibinfo{volume}{15}
  (\bibinfo{date}{Feb.} \bibinfo{year}{2022}), \bibinfo{pages}{796895}.
\newblock


\bibitem[Ling et~al\mbox{.}(2003)]%
        {auc}
\bibfield{author}{\bibinfo{person}{Charles~X. Ling}, \bibinfo{person}{Jin
  Huang}, {and} \bibinfo{person}{Harry Zhang}.}
  \bibinfo{year}{2003}\natexlab{}.
\newblock \showarticletitle{AUC: a statistically consistent and more
  discriminating measure than accuracy}. In
  \bibinfo{booktitle}{\emph{Proceedings of the 18th International Joint
  Conference on Artificial Intelligence}} (Acapulco, Mexico)
  \emph{(\bibinfo{series}{IJCAI'03})}. \bibinfo{publisher}{Morgan Kaufmann
  Publishers Inc.}, \bibinfo{address}{San Francisco, CA, USA},
  \bibinfo{pages}{519–524}.
\newblock


\bibitem[Morrison et~al\mbox{.}(2024)]%
        {incorrect_xai_1}
\bibfield{author}{\bibinfo{person}{Katelyn Morrison}, \bibinfo{person}{Philipp
  Spitzer}, \bibinfo{person}{Violet Turri}, \bibinfo{person}{Michelle Feng},
  \bibinfo{person}{Niklas K\"{u}hl}, {and} \bibinfo{person}{Adam Perer}.}
  \bibinfo{year}{2024}\natexlab{}.
\newblock \showarticletitle{The Impact of Imperfect XAI on Human-AI
  Decision-Making}.
\newblock \bibinfo{journal}{\emph{Proc. ACM Hum.-Comput. Interact.}}
  \bibinfo{volume}{8}, \bibinfo{number}{CSCW1}, Article
  \bibinfo{articleno}{183} (\bibinfo{date}{April} \bibinfo{year}{2024}),
  \bibinfo{numpages}{39}~pages.
\newblock
\urldef\tempurl%
\url{https://doi.org/10.1145/3641022}
\showDOI{\tempurl}


\bibitem[Ouwehand et~al\mbox{.}(2021)]%
        {cognitive_load_rating}
\bibfield{author}{\bibinfo{person}{Kim Ouwehand}, \bibinfo{person}{Avalon
  van~der Kroef}, \bibinfo{person}{Jacqueline Wong}, {and}
  \bibinfo{person}{Fred Paas}.} \bibinfo{year}{2021}\natexlab{}.
\newblock \showarticletitle{Measuring Cognitive Load: Are There More Valid
  Alternatives to Likert Rating Scales?}
\newblock \bibinfo{journal}{\emph{Frontiers in Education}}
  \bibinfo{volume}{Volume 6 - 2021} (\bibinfo{year}{2021}).
\newblock
\urldef\tempurl%
\url{https://www.frontiersin.org/journals/education/articles/10.3389/feduc.2021.702616,
  DOI={10.3389/feduc.2021.702616}, ISSN={2504-284X},}
\showURL{%
\tempurl}


\bibitem[Parikh(2018)]%
        {Parikh2018EyeGF}
\bibfield{author}{\bibinfo{person}{Saurin~S. Parikh}.}
  \bibinfo{year}{2018}\natexlab{}.
\newblock \showarticletitle{Eye Gaze Feature Classification for Predicting
  Levels of Learning}.
\newblock
\urldef\tempurl%
\url{https://api.semanticscholar.org/CorpusID:53471366}
\showURL{%
\tempurl}


\bibitem[Peirce et~al\mbox{.}(2019)]%
        {psychopy}
\bibfield{author}{\bibinfo{person}{Jonathan Peirce}, \bibinfo{person}{Jeremy~R
  Gray}, \bibinfo{person}{Sol Simpson}, \bibinfo{person}{Michael MacAskill},
  \bibinfo{person}{Richard H{\"o}chenberger}, \bibinfo{person}{Hiroyuki Sogo},
  \bibinfo{person}{Erik Kastman}, {and} \bibinfo{person}{Jonas~Kristoffer
  Lindel{\o}v}.} \bibinfo{year}{2019}\natexlab{}.
\newblock \showarticletitle{{PsychoPy2}: Experiments in behavior made easy}.
\newblock \bibinfo{journal}{\emph{Behavior Research Methods}}
  \bibinfo{volume}{51}, \bibinfo{number}{1} (\bibinfo{date}{Feb.}
  \bibinfo{year}{2019}), \bibinfo{pages}{195--203}.
\newblock


\bibitem[Romeo and Conti(2025)]%
        {bias_ai}
\bibfield{author}{\bibinfo{person}{Giuseppe Romeo} {and}
  \bibinfo{person}{Daniela Conti}.} \bibinfo{year}{2025}\natexlab{}.
\newblock \showarticletitle{Exploring automation bias in {human--AI}
  collaboration: a review and implications for explainable {AI}}.
\newblock \bibinfo{journal}{\emph{AI Soc.}} (\bibinfo{date}{July}
  \bibinfo{year}{2025}).
\newblock


\bibitem[Ross and Willson(2017)]%
        {t-test}
\bibfield{author}{\bibinfo{person}{Amanda Ross} {and}
  \bibinfo{person}{Victor~L. Willson}.} \bibinfo{year}{2017}\natexlab{}.
\newblock \bibinfo{booktitle}{\emph{Paired Samples T-Test}}.
\newblock \bibinfo{publisher}{SensePublishers}, \bibinfo{address}{Rotterdam},
  \bibinfo{pages}{17--19}.
\newblock
\showISBNx{978-94-6351-086-8}
\urldef\tempurl%
\url{https://doi.org/10.1007/978-94-6351-086-8_4}
\showDOI{\tempurl}


\bibitem[Salvucci and Goldberg(2000)]%
        {tobii_filter}
\bibfield{author}{\bibinfo{person}{Dario~D. Salvucci} {and}
  \bibinfo{person}{Joseph~H. Goldberg}.} \bibinfo{year}{2000}\natexlab{}.
\newblock \showarticletitle{Identifying fixations and saccades in eye-tracking
  protocols}. In \bibinfo{booktitle}{\emph{Proceedings of the 2000 Symposium on
  Eye Tracking Research \& Applications}} (Palm Beach Gardens, Florida, USA)
  \emph{(\bibinfo{series}{ETRA '00})}. \bibinfo{publisher}{Association for
  Computing Machinery}, \bibinfo{address}{New York, NY, USA},
  \bibinfo{pages}{71–78}.
\newblock
\showISBNx{1581132808}
\urldef\tempurl%
\url{https://doi.org/10.1145/355017.355028}
\showDOI{\tempurl}


\bibitem[Schemmer et~al\mbox{.}(2023)]%
        {appropriate_reliance_AI_advic}
\bibfield{author}{\bibinfo{person}{Max Schemmer}, \bibinfo{person}{Niklas
  Kuehl}, \bibinfo{person}{Carina Benz}, \bibinfo{person}{Andrea Bartos}, {and}
  \bibinfo{person}{Gerhard Satzger}.} \bibinfo{year}{2023}\natexlab{}.
\newblock \showarticletitle{Appropriate Reliance on AI Advice:
  Conceptualization and the Effect of Explanations}. In
  \bibinfo{booktitle}{\emph{Proceedings of the 28th International Conference on
  Intelligent User Interfaces}} (Sydney, NSW, Australia)
  \emph{(\bibinfo{series}{IUI '23})}. \bibinfo{publisher}{Association for
  Computing Machinery}, \bibinfo{address}{New York, NY, USA},
  \bibinfo{pages}{410–422}.
\newblock
\showISBNx{9798400701061}
\urldef\tempurl%
\url{https://doi.org/10.1145/3581641.3584066}
\showDOI{\tempurl}


\bibitem[Son(2016)]%
        {cold_start}
\bibfield{author}{\bibinfo{person}{Le~Hoang Son}.}
  \bibinfo{year}{2016}\natexlab{}.
\newblock \showarticletitle{Dealing with the new user cold-start problem in
  recommender systems: A comparative review}.
\newblock \bibinfo{journal}{\emph{Information Systems}}  \bibinfo{volume}{58}
  (\bibinfo{year}{2016}), \bibinfo{pages}{87--104}.
\newblock
\showISSN{0306-4379}
\urldef\tempurl%
\url{https://doi.org/10.1016/j.is.2014.10.001}
\showDOI{\tempurl}


\bibitem[Springer and Whittaker(2020)]%
        {Springer2020}
\bibfield{author}{\bibinfo{person}{Aaron Springer} {and} \bibinfo{person}{Steve
  Whittaker}.} \bibinfo{year}{2020}\natexlab{}.
\newblock \showarticletitle{Progressive Disclosure: When, Why, and How Do Users
  Want Algorithmic Transparency Information?}
\newblock \bibinfo{journal}{\emph{ACM Trans. Interact. Intell. Syst.}}
  \bibinfo{volume}{10}, \bibinfo{number}{4}, Article \bibinfo{articleno}{29}
  (\bibinfo{date}{Oct.} \bibinfo{year}{2020}), \bibinfo{numpages}{32}~pages.
\newblock
\showISSN{2160-6455}
\urldef\tempurl%
\url{https://doi.org/10.1145/3374218}
\showDOI{\tempurl}


\bibitem[Sun et~al\mbox{.}(2026a)]%
        {sun_trust}
\bibfield{author}{\bibinfo{person}{Xin Sun}, \bibinfo{person}{Rongjun Ma},
  \bibinfo{person}{Shu Wei}, \bibinfo{person}{Pablo Cesar},
  \bibinfo{person}{Jos~A. Bosch}, {and} \bibinfo{person}{Abdallah {El Ali}}.}
  \bibinfo{year}{2026}\natexlab{a}.
\newblock \showarticletitle{Understanding trust toward human versus
  AI-generated health information through behavioral and physiological
  sensing}.
\newblock \bibinfo{journal}{\emph{International Journal of Human-Computer
  Studies}}  \bibinfo{volume}{209} (\bibinfo{year}{2026}),
  \bibinfo{pages}{103714}.
\newblock
\showISSN{1071-5819}
\urldef\tempurl%
\url{https://doi.org/10.1016/j.ijhcs.2025.103714}
\showDOI{\tempurl}


\bibitem[Sun et~al\mbox{.}(2026b)]%
        {xsun_reasoning_chi26}
\bibfield{author}{\bibinfo{person}{Xin Sun}, \bibinfo{person}{Shu Wei},
  \bibinfo{person}{Jos~A Bosch}, \bibinfo{person}{Isao Echizen},
  \bibinfo{person}{Saku Sugawara}, {and} \bibinfo{person}{Abdallah El~Ali}.}
  \bibinfo{year}{2026}\natexlab{b}.
\newblock \showarticletitle{Seeing the Reasoning: How LLM Rationales Influence
  User Trust and Decision-Making in Factual Verification Tasks}. In
  \bibinfo{booktitle}{\emph{Proceedings of the Extended Abstracts of the 2026
  CHI Conference on Human Factors in Computing Systems}}
  \emph{(\bibinfo{series}{CHI EA '26})}. \bibinfo{publisher}{Association for
  Computing Machinery}, \bibinfo{address}{New York, NY, USA}, Article
  \bibinfo{articleno}{585}, \bibinfo{numpages}{7}~pages.
\newblock
\showISBNx{9798400722813}
\urldef\tempurl%
\url{https://doi.org/10.1145/3772363.3798613}
\showDOI{\tempurl}


\bibitem[Taudien et~al\mbox{.}(2022)]%
        {confidence_effect_in_decision_2}
\bibfield{author}{\bibinfo{person}{Anna Taudien}, \bibinfo{person}{Andreas
  Fügener}, \bibinfo{person}{Alok Gupta}, {and} \bibinfo{person}{Wolfgang
  Ketter}.} \bibinfo{year}{2022}\natexlab{}.
\newblock \showarticletitle{The Effect of AI Advice on Human Confidence in
  Decision-Making}.
\newblock
\urldef\tempurl%
\url{https://doi.org/10.24251/HICSS.2022.029}
\showDOI{\tempurl}


\bibitem[{Tobii AB}(2024)]%
        {tobii_pro}
\bibfield{author}{\bibinfo{person}{{Tobii AB}}.}
  \bibinfo{year}{2024}\natexlab{}.
\newblock \bibinfo{title}{{Tobii Pro Lab (Version 1.2xx)}}.
\newblock
\newblock


\bibitem[Toker and Conati(2017)]%
        {cognitive_pupil}
\bibfield{author}{\bibinfo{person}{Dereck Toker} {and}
  \bibinfo{person}{Cristina Conati}.} \bibinfo{year}{2017}\natexlab{}.
\newblock \showarticletitle{Leveraging Pupil Dilation Measures for
  Understanding Users' Cognitive Load During Visualization Processing}. In
  \bibinfo{booktitle}{\emph{Adjunct Publication of the 25th Conference on User
  Modeling, Adaptation and Personalization}} (Bratislava, Slovakia)
  \emph{(\bibinfo{series}{UMAP '17})}. \bibinfo{publisher}{Association for
  Computing Machinery}, \bibinfo{address}{New York, NY, USA},
  \bibinfo{pages}{267–270}.
\newblock
\showISBNx{9781450350679}
\urldef\tempurl%
\url{https://doi.org/10.1145/3099023.3099059}
\showDOI{\tempurl}


\bibitem[Van~der Lans et~al\mbox{.}(2011)]%
        {fixations}
\bibfield{author}{\bibinfo{person}{Ralf Van~der Lans}, \bibinfo{person}{Michel
  Wedel}, {and} \bibinfo{person}{Rik Pieters}.}
  \bibinfo{year}{2011}\natexlab{}.
\newblock \showarticletitle{Defining eye-fixation sequences across individuals
  and tasks: the Binocular-Individual Threshold (BIT) algorithm}.
\newblock \bibinfo{journal}{\emph{Behavior research methods}}
  \bibinfo{volume}{43} (\bibinfo{year}{2011}), \bibinfo{pages}{239--257}.
\newblock


\bibitem[Vasconcelos et~al\mbox{.}(2023)]%
        {explanation_ai_overreliance}
\bibfield{author}{\bibinfo{person}{Helena Vasconcelos},
  \bibinfo{person}{Matthew Jörke}, \bibinfo{person}{Madeleine
  Grunde-McLaughlin}, \bibinfo{person}{Tobias Gerstenberg},
  \bibinfo{person}{Michael Bernstein}, {and} \bibinfo{person}{Ranjay Krishna}.}
  \bibinfo{year}{2023}\natexlab{}.
\newblock \bibinfo{title}{Explanations Can Reduce Overreliance on AI Systems
  During Decision-Making}.
\newblock
\newblock
\showeprint[arxiv]{2212.06823}~[cs.HC]
\urldef\tempurl%
\url{https://arxiv.org/abs/2212.06823}
\showURL{%
\tempurl}


\bibitem[Vereschak et~al\mbox{.}(2024)]%
        {Vereschak2024}
\bibfield{author}{\bibinfo{person}{Oleksandra Vereschak},
  \bibinfo{person}{Fatemeh Alizadeh}, \bibinfo{person}{Gilles Bailly}, {and}
  \bibinfo{person}{Baptiste Caramiaux}.} \bibinfo{year}{2024}\natexlab{}.
\newblock \showarticletitle{Trust in AI-assisted Decision Making: Perspectives
  from Those Behind the System and Those for Whom the Decision is Made}. In
  \bibinfo{booktitle}{\emph{Proceedings of the CHI Conference on Human Factors
  in Computing Systems}} (Honolulu, HI, USA) \emph{(\bibinfo{series}{CHI
  '24})}. \bibinfo{publisher}{Association for Computing Machinery},
  \bibinfo{address}{New York, NY, USA}, Article \bibinfo{articleno}{28},
  \bibinfo{numpages}{14}~pages.
\newblock
\showISBNx{9798400703300}
\urldef\tempurl%
\url{https://doi.org/10.1145/3613904.3642018}
\showDOI{\tempurl}


\bibitem[Vereschak et~al\mbox{.}(2021)]%
        {confidence_rating}
\bibfield{author}{\bibinfo{person}{Oleksandra Vereschak},
  \bibinfo{person}{Gilles Bailly}, {and} \bibinfo{person}{Baptiste Caramiaux}.}
  \bibinfo{year}{2021}\natexlab{}.
\newblock \showarticletitle{How to Evaluate Trust in AI-Assisted Decision
  Making? A Survey of Empirical Methodologies}.
\newblock \bibinfo{journal}{\emph{Proc. ACM Hum.-Comput. Interact.}}
  \bibinfo{volume}{5}, \bibinfo{number}{CSCW2}, Article
  \bibinfo{articleno}{327} (\bibinfo{date}{Oct.} \bibinfo{year}{2021}),
  \bibinfo{numpages}{39}~pages.
\newblock
\urldef\tempurl%
\url{https://doi.org/10.1145/3476068}
\showDOI{\tempurl}


\bibitem[Wang and Stern(2001)]%
        {saccades}
\bibfield{author}{\bibinfo{person}{L Wang} {and} \bibinfo{person}{J~A Stern}.}
  \bibinfo{year}{2001}\natexlab{}.
\newblock \showarticletitle{Saccade initiation and accuracy in gaze shifts are
  affected by visual stimulus significance}.
\newblock \bibinfo{journal}{\emph{Psychophysiology}} \bibinfo{volume}{38},
  \bibinfo{number}{1} (\bibinfo{date}{Jan.} \bibinfo{year}{2001}),
  \bibinfo{pages}{64--75}.
\newblock


\bibitem[Wang et~al\mbox{.}(2016)]%
        {7451741}
\bibfield{author}{\bibinfo{person}{Ning Wang}, \bibinfo{person}{David~V.
  Pynadath}, {and} \bibinfo{person}{Susan~G. Hill}.}
  \bibinfo{year}{2016}\natexlab{}.
\newblock \showarticletitle{Trust calibration within a human-robot team:
  Comparing automatically generated explanations}. In
  \bibinfo{booktitle}{\emph{2016 11th ACM/IEEE International Conference on
  Human-Robot Interaction (HRI)}}. \bibinfo{pages}{109--116}.
\newblock
\urldef\tempurl%
\url{https://doi.org/10.1109/HRI.2016.7451741}
\showDOI{\tempurl}


\bibitem[Wang et~al\mbox{.}(2023)]%
        {Wang2023-nm}
\bibfield{author}{\bibinfo{person}{Xujinfeng Wang}, \bibinfo{person}{Yicheng
  Yang}, \bibinfo{person}{Da Tao}, {and} \bibinfo{person}{Tingru Zhang}.}
  \bibinfo{year}{2023}\natexlab{}.
\newblock \showarticletitle{The impact of AI transparency and reliability on
  human-AI collaborative decision-making}. In \bibinfo{booktitle}{\emph{{AHFE}
  International}}. \bibinfo{publisher}{AHFE International}.
\newblock


\bibitem[Wedel et~al\mbox{.}(2023)]%
        {eye_movements_decision}
\bibfield{author}{\bibinfo{person}{Michel Wedel}, \bibinfo{person}{Rik
  Pieters}, {and} \bibinfo{person}{Ralf van~der Lans}.}
  \bibinfo{year}{2023}\natexlab{}.
\newblock \showarticletitle{Modeling eye movements during decision making: A
  review}.
\newblock \bibinfo{journal}{\emph{Psychometrika}} \bibinfo{volume}{88},
  \bibinfo{number}{2} (\bibinfo{date}{June} \bibinfo{year}{2023}),
  \bibinfo{pages}{697--729}.
\newblock


\bibitem[Wikipwdia(2023)]%
        {Hallucin28:online}
\bibfield{author}{\bibinfo{person}{Wikipwdia}.}
  \bibinfo{year}{2023}\natexlab{}.
\newblock \bibinfo{title}{Hallucination (artificial intelligence) - Wikipedia}.
\newblock
  \bibinfo{howpublished}{\url{https://en.wikipedia.org/wiki/Hallucination_(artificial_intelligence)}}.
\newblock
\newblock
\shownote{(Accessed on 10/10/2023)}.


\bibitem[Yao et~al\mbox{.}(2025)]%
        {yao2025factcheckingaigeneratednewsreports}
\bibfield{author}{\bibinfo{person}{Jiayi Yao}, \bibinfo{person}{Haibo Sun},
  {and} \bibinfo{person}{Nianwen Xue}.} \bibinfo{year}{2025}\natexlab{}.
\newblock \bibinfo{title}{Fact-checking AI-generated news reports: Can LLMs
  catch their own lies?}
\newblock
\newblock
\showeprint[arxiv]{2503.18293}~[cs.CL]
\urldef\tempurl%
\url{https://arxiv.org/abs/2503.18293}
\showURL{%
\tempurl}


\bibitem[Zhang et~al\mbox{.}(2020)]%
        {confidence_effect_in_decision_1}
\bibfield{author}{\bibinfo{person}{Yunfeng Zhang}, \bibinfo{person}{Q.~Vera
  Liao}, {and} \bibinfo{person}{Rachel K.~E. Bellamy}.}
  \bibinfo{year}{2020}\natexlab{}.
\newblock \showarticletitle{Effect of Confidence and Explanation on Accuracy
  and Trust Calibration in AI-Assisted Decision Making}
  \emph{(\bibinfo{series}{FAT* '20})}. \bibinfo{publisher}{ACM},
  \bibinfo{address}{New York, NY, USA}, \bibinfo{pages}{295–305}.
\newblock
\showISBNx{9781450369367}
\urldef\tempurl%
\url{https://doi.org/10.1145/3351095.3372852}
\showDOI{\tempurl}


\bibitem[Ömer Sümer et~al\mbox{.}(2021)]%
        {SUMER2021106909}
\bibfield{author}{\bibinfo{person}{Ömer Sümer}, \bibinfo{person}{Efe Bozkir},
  \bibinfo{person}{Thomas Kübler}, \bibinfo{person}{Sven Grüner},
  \bibinfo{person}{Sonja Utz}, {and} \bibinfo{person}{Enkelejda Kasneci}.}
  \bibinfo{year}{2021}\natexlab{}.
\newblock \showarticletitle{FakeNewsPerception: An eye movement dataset on the
  perceived believability of news stories}.
\newblock \bibinfo{journal}{\emph{Data in Brief}}  \bibinfo{volume}{35}
  (\bibinfo{year}{2021}), \bibinfo{pages}{106909}.
\newblock
\showISSN{2352-3409}
\urldef\tempurl%
\url{https://doi.org/10.1016/j.dib.2021.106909}
\showDOI{\tempurl}


\bibitem[Štěpán Novák et~al\mbox{.}(2024)]%
        {eyetracking_survey}
\bibfield{author}{\bibinfo{person}{Jakub Štěpán Novák},
  \bibinfo{person}{Jan Masner}, \bibinfo{person}{Petr Benda},
  \bibinfo{person}{Pavel Šimek}, {and} \bibinfo{person}{Vojtěch Merunka}.}
  \bibinfo{year}{2024}\natexlab{}.
\newblock \showarticletitle{Eye Tracking, Usability, and User Experience: A
  Systematic Review}.
\newblock \bibinfo{journal}{\emph{International Journal of Human–Computer
  Interaction}} \bibinfo{volume}{40}, \bibinfo{number}{17}
  (\bibinfo{year}{2024}), \bibinfo{pages}{4484--4500}.
\newblock
\urldef\tempurl%
\url{https://doi.org/10.1080/10447318.2023.2221600}
\showDOI{\tempurl}
\showeprint{https://doi.org/10.1080/10447318.2023.2221600}


\end{thebibliography}

\end{document}